\documentclass[fleqn,usenatbib]{mnras}

\usepackage{caption}
\usepackage{amsmath}
\usepackage{amssymb}
\usepackage{multicol}
\usepackage{rotating}
\usepackage{lscape}
\usepackage{graphicx}
\usepackage{color}
\usepackage{placeins}
\usepackage{xcolor}
\usepackage{float}
\usepackage{verbatim}
\usepackage{bbold}
\usepackage{colortbl} 
\usepackage{lipsum,adjustbox}
\usepackage{multirow}

\usepackage{tikz}
\usetikzlibrary{matrix, fit, positioning,arrows.meta,arrows, ,decorations.pathreplacing,shapes.geometric,backgrounds}

\tikzstyle{startstop}=[rectangle, rounded corners, minimum width=2.7cm, minimum height= 1cm, text centered, draw=black, fill=red!20, text width = 2cm]
\tikzstyle{startstop_nb}=[rectangle, rounded corners, minimum width=2.5cm, minimum height=0.8cm, text centered, draw=black, fill=gray!20, text width = 2cm]
\tikzstyle{io}=[trapezium, trapezium left angle = 70, trapezium right angle =110,minimum width=2cm, minimum height= 1cm, text centered, draw=black, fill=blue!30]
\tikzstyle{io_short}=[rectangle,rounded corners, minimum width=0.2cm, minimum height= 1cm, text centered, draw=black, fill=blue!5, line width = 0.05cm]
\tikzstyle{process}=[rectangle, rounded corners,minimum width=3cm, minimum height= 0.7cm, text centered, draw=black, fill=orange!30]
\tikzstyle{process_DUMMY2}=[rectangle, minimum width=2.7cm, minimum height= 1cm, text centered, draw=white, fill=white!0, text width = 3cm]
\tikzstyle{process_DUMMY}=[rectangle, minimum width=0cm, minimum height= 0cm, draw=white, fill=white!0]
\tikzstyle{process_SDUMMY}=[circle, minimum width=0cm, minimum height= 0cm, draw=white, fill=white!0]

\tikzstyle{process_NO_BOX}=[circle, minimum width=0.5cm, minimum height= 0.5cm, text centered, draw=gray, fill=white!0]
\tikzstyle{startstop3}=[rectangle, rounded corners, minimum width=2.7cm, minimum height= 1cm, text centered, draw=black, fill=white, text width = 3cm]

\tikzstyle{process_TITLE}=[rectangle, rounded corners, minimum width=1cm, minimum height= 0.4cm, text centered, draw=black, fill=gray!30]
\tikzstyle{nomen}=[rectangle, rounded corners, minimum width=4cm, minimum height= 1cm,  draw=black, fill=yellow!10]
\tikzstyle{process_short}=[rectangle, rounded corners,minimum width=1cm, minimum height=0.5cm, text centered, draw=black, fill=orange!20]
\tikzstyle{decision}=[diamond, rounded corners,minimum width=2cm, minimum height= 0.7cm, text centered, draw=black, fill=green!20]
\tikzstyle{question}=[diamond,rounded corners, minimum width=2.5cm, minimum height= 0.7cm, text centered, draw=black, fill=blue!20]
\tikzstyle{results}=[rectangle,rounded corners, minimum width=2.5cm, minimum height= 0.7cm, text centered, draw=black, fill=blue!20]
\tikzstyle{decisionfwd}=[diamond, rounded corners,minimum width=2cm, minimum height= 0.7cm, text centered, draw=black, fill=green!20]
\tikzstyle{decision_short}=[diamond, rounded corners,minimum width=1cm, minimum height= 1cm, text centered, draw=black, fill=green!20]
\tikzstyle{decision_short_input}=[diamond, minimum width=0.8cm, minimum height=0.05cm, text centered, draw=black, fill=gray!20]
\tikzstyle{arrow}=[thick, ->, >= stealth]
\tikzstyle{arrow_d}=[dotted, ->, >= stealth]
\tikzstyle{arrow_new}=[dotted, ->, >= stealth]
\tikzstyle{arrow_pi}=[dashed, ->, >= stealth, line width=0.03cm]
\tikzstyle{arrow_nl}=[dashed, -, >= stealth, line width=0.03cm]
\tikzstyle{arrow_nls}=[thick, -, >= stealth, line width=0.03cm]
\tikzstyle{arrow_new}=[dotted, ->,>= stealth, color=red, line width = 0.02cm]
\tikzstyle{arrow_pib}=[dashed, ->, >= stealth, line width=0.03cm, color=blue]
\tikzstyle{arrow_nlb}=[dashed, -, >= stealth, line width=0.03cm, color=blue]

\newcommand{\beq}{\begin{equation}}
\newcommand{\eeq}{\end{equation}}
\newcommand{\beqa}{\begin{eqnarray}}
\newcommand{\eeqa}{\end{eqnarray}}

\definecolor{gray}{gray}{0.55}

\font\BF=cmmib10

\def\v{{\hbox{\BF v}}}

\newcommand{\kmsmpc}{\mbox{km\,s$^{-1}$\,Mpc$^{-1}$}}
\newcommand{\mbi}[1]{\mbox{\boldmath$#1$}}

\newcommand{\mat}[1]{\mbox{\rm\bf #1}}
\newcommand{\lsim}{\mbox{${\,\hbox{\hbox{$ < $}\kern -0.8em \lower 1.0ex\hbox{$\sim$}}\,}$}}
\newcommand{\gsim}{\mbox{${\,\hbox{\hbox{$ > $}\kern -0.8em \lower 1.0ex\hbox{$\sim$}}\,}$}}

\def\beqn{\vspace{2mm}
\begin{eqnarray}}
\def\eeqn{\vspace{2mm}
\end{eqnarray}}

 \def\la{\langle}

\newcommand{\be}{\begin{equation}}
\newcommand{\ee}{\end{equation}}
\newcommand{\ba}{\begin{eqnarray}}
\newcommand{\ea}{\end{eqnarray}}
\newcommand{\brr}{\begin{array}}

\newcommand{\err}{\end{array}}
\newcommand{\bc}{\begin{center}}
\newcommand{\ec}{\end{center}}

\title[Higher Order HMC]{Higher Order Hamiltonian Monte Carlo Sampling for Cosmological Large-Scale Structure Analysis}

\author[Hern{\'a}ndez-S{\'a}nchez, Kitaura, Ata \& Dalla Vecchia]{M{\'o}nica Hern{\'a}ndez-S{\'a}nchez$^{1,2}$\thanks{E-mail:\href{mailto:mhs@iac.es}{mhs@iac.es}}, Francisco-Shu~Kitaura$^{1,2}$\thanks{E-mail:\href{mailto:fkitaura@iac.es}{fkitaura@iac.es}},  Metin Ata$^{3}$\thanks{E-mail:\href{mailto:metinata.mail@gmail.com}{metin.ata@ipmu.jp}}, \and and Claudio Dalla Vecchia$^{1,2}$ 
\\ \\
$^1$Instituto de Astrof\'{\i}sica de Canarias (IAC), Calle V\'{\i}a L{\'a}ctea s/n, 38200, La Laguna, Tenerife, Spain \\ 
$^2$Departamento de Astrof\'{\i}sica, Universidad de La Laguna (ULL), E-38206, La Laguna, Tenerife, Spain\\
$^3$Kavli Institute for the Physics and Mathematics of the Universe (Kavli IPMU), WPI,\\   UTIAS, The University of Tokyo, Kashiwa, Chiba, 277-8568, Japan}

\date{Accepted XXX. Received YYY; in original form ZZZ}
\pubyear{2019}

\begin{document}
\label{firstpage}
\pagerange{\pageref{firstpage}--\pageref{lastpage}}
\maketitle

\begin{abstract} 
{We investigate higher order symplectic integration strategies within Bayesian cosmic density field reconstruction methods. In particular, we study the fourth-order discretisation of Hamiltonian equations of motion (EoM). This is achieved by recursively applying the basic second-order leap-frog scheme (considering the single evaluation of the EoM) in a combination of  even numbers of forward time integration steps with a single intermediate backward step. This largely reduces the number of evaluations and random gradient computations, as required in the usual second-order case for high-dimensional cases. We restrict this study to the lognormal-Poisson model, applied to a full volume halo catalogue in real space on a cubical  mesh of 1250 $h^{-1}$ Mpc side and 256$^3$ cells. Hence, we neglect selection effects, redshift space distortions, and displacements. We note that those observational and cosmic evolution effects can be accounted for in subsequent Gibbs-sampling steps within the \texttt{COSMIC BIRTH} algorithm. We find that going from the usual second to fourth-order in the leap-frog scheme  shortens the burn-in phase by a factor of at least $\sim30$.
This implies that 75-90 independent samples are obtained while the fastest second-order method converges.  
After convergence, the correlation lengths indicate an improvement factor of about {\color{black}$3.0$} fewer gradient computations  {\color{black}for meshes of 256$^3$ cells}.  {\color{black}In the considered cosmological scenario, the traditional leap-frog scheme turns out to outperform higher order integration schemes only when considering lower dimensional problems, e.g. meshes with 64$^3$ cells. } This gain in computational efficiency  can help to go towards a full Bayesian analysis of the cosmological large-scale structure for upcoming galaxy surveys.}
\end{abstract}

\begin{keywords}
galaxies: distances and redshifts -- large-scale structure of Universe -- methods: statistical -- methods: analytical -- cosmology: observations
\end{keywords}




\section{Introduction}

In the current cosmological picture, the non-linear structures we observe today have risen from some closely Gaussian primordial fluctuations \citep[see e.g.][and references therein]{2010gfe..book.....M}.
Gaussian fields have the convenient property of being fully characterised by the variance, i.e. the two-point statistics, which is given by the correlation function in configuration space, or the power spectrum in Fourier space.
It is thus common to extract cosmological information from the two-point statistics \citep[see e.g.][]{2017MNRAS.471.2370C,2017MNRAS.464.3409B,2017MNRAS.464.1168R}. However, as gravity couples different scales, the cosmic density field is far from being Gaussian anymore, and the linear predictions of the two-point statistics do not match the observations \citep[see e.g.][]{2018MNRAS.473.1195L}. Therefore, non-linear models have been developed to be able to compare the theoretical predictions to the observations and constrain cosmological parameters \citep[see e.g.][]{2008MNRAS.383..755A,2009PASJ...61..321N,2011MNRAS.417.1913R,2015PhRvD..92j3516O,2015MNRAS.450.3822W,2015PhRvD..91h4010U,2017PhRvD..96d3526H,2017JCAP...08..029B}.
But even if one succeeds in doing so, not all
the cosmological information is encoded in the two-point statistics in low redshift data, as opposed to the cosmic microwave background \citep[see e.g.][]{2015PhRvD..92l3522S}. This is why linearisation methods have been suggested in the literature \citep[][]{2009ApJ...698L..90N,2012MNRAS.425.2443K,2016MNRAS.459.1916S}. In particular, reconstruction takes the galaxies back in time, putting back information from the higher order into the two-point statistics, thus increasing the precision of baryon acoustic oscillation (BAO) signature measurement \citep{2007ApJ...664..675E,2012MNRAS.427.2132P}. Other ways of gaining non-linear information from the galaxy distribution  have been suggested based on  the three-point statistics  \citep[see e.g.][]{2014PhRvD..90l3522S,2017MNRAS.465.1757G}, or on reconstructions of  cosmic voids \citep{2016PhRvL.116q1301K,2018arXiv180203990Z}.
From a Bayesian perspective, one can write the posterior distribution function relating the primordial density field to the galaxy distribution through a Gaussian prior and some likelihood including non-linear dynamics and some bias description \citep{argo}. 
The resulting global posterior probability distribution function (PDF) is clearly non-Gaussian. 
One of the simplest models we can consider is the lognormal-Poisson, accounting for the non-Gaussian matter distribution and the discreteness of the galaxy distribution \citep{argo3}.
More complex variations on this can be suggested, including deviations from Poissonity in the likelihood, or non-linear dynamics in the connection between the initial and final cosmic density field.
As a matter of fact, the lognormal-Poisson model can be an accurate model for Lagrangian tracers, which are connected within a Gibbs-sampling scheme  to the observed galaxy field distribution sampling the displacements in a separated step \citep{2019arXiv191100284K}. 
Hamiltonian Monte Carlo techniques permit us to sample from non-Gaussian PDFs \citep{DUANE1987216,Neal,fasthamiltonian,MCMC}. Ever since the first application to observational data from galaxy surveys without \citep{2010MNRAS.409..355J} and with cosmic evolution modelling \citep{2012MNRAS.427L..35K}, a number of  Bayesian inference methods have been developed to solve the problem of sampling linear density fields from a galaxy distribution \citep{2013MNRAS.432..894J,Kitaura2013,2013ApJ...772...63W,2014ApJ...794...94W,2019MNRAS.488.2573B,2019A&A...625A..64J}. However, one of the drawbacks of these methods is that they require thousands of accepted iterations until convergence, and have very long correlation lengths of several hundred to one thousand iterations. As galaxy surveys increase in volume, accurate reconstructions demand meshes with between one hundred million to one billion cells. Given the high dimensionality of the problem, Bayesian methods cannot be considered yet to be practical to sample full posterior distributions and constrain cosmological parameters.

This calls for efforts in increasing the efficiency of the Hamiltonian Monte Carlo Sampling. 
A number of works have investigated higher order discretisations of the Hamiltonian equations of motions
\citep[for a comprehensive summary see][]{hairer_lubich_wanner_2010}. \citet{YOSHIDA1990262} proposed a higher order symplectic integration parametrising the integration steps and calculating the exact coefficients.   
Also, efforts have been  done in the field of quantum-chromodynamics and lattice computations, successively applying second-order leap-frog integrations \citep{1988PhRvD..38.1228C,creutz,CAMPOSTRINI1990753,Kennedy:2006ax,Luscher:2010ae}. In the field of applied mathematics, \citet{Blanes_2014} suggested a higher order integrator by sampling from Gaussian distributions and splitting the integration scheme, evaluating the force term several times per integration step. This has been incorporated into a general $N$-body integration framework in \citet{2018MNRAS.473.3351R}. Multisymplectic integrators  \citep[e.g.][]{ISLAS2004585} are extensively used describing the evolution of the Schr\"{o}dinger equation in quantum field theory. For other advances in higher order symplectic integration methods see \citet{Omelayan}. 

{\color{black} Also other works have investigated higher order discretisation schemes \citep[][]{mannseth2016application,aless2017geometry,pmlr-v37-chao15}, in particular in the field of the cosmic microwave background  \citep[][]{Taylor_2008,Souradeep_2016}, however, without aiming at accelerating the Hamiltonian sampling method, or with very little success in this aspect. }

In this work, we investigate the computational efficiency of the fourth-order leap-frog scheme {\color{black} by recursively applying the basic second-order leap-frog scheme in a combination of  even numbers of forward time integration steps with a single intermediate backward step, following the works of \citep{1988PhRvD..38.1228C,creutz,CAMPOSTRINI1990753}. It is important to stress, that, instead of applying the second-order leap-frog scheme as it is usually done within the recursive formula, we consider only the single evaluation of the Hamiltonian equations of motion. This means that we do not include the randomization of the number of steps for fourth-order scheme. In this way, we can squeeze the potential of the larger integration steps allowed by the higher order scheme, as we show in this paper.   }  
{\color{black} Recent studies indicate that clever applications of recursive forward and backward second-order leap-frog computations can yield significant  efficiency improvements in high dimensional spaces \citep[see the NUTS scheme][showing improvements of up to factors of 3]{2011arXiv1111.4246H}. }
Generally, higher accurate symplectic integrators are computationally more expensive, thus it is important to find a beneficial trade-off for computational costs and gains in phase-space movement. We show in this work that higher order integrators can become significantly more efficient in high statistical dimensions, as expected from mathematical considerations \citep[see][]{2010arXiv1001.4460B,2017arXiv171105337B}.
{\color{black} For higher order methods solving classical Hamiltonian systems see  \citet{2002McLachlan}. This field of research indicates that more sophisticated higher order schemes than the one studied in this paper are very promising for Bayesian studies  \citep[see][]{2002McLachlanQuispel}.}

The work presented here potentially represents a major step forward in Bayesian inference studies within cosmological large-scale structure analysis.
This paper is a companion paper of  \citet{2019arXiv191100284K}.

The remainder of this manuscript is structured as follows, first we revise the theory of Hamiltonian Monte Carlo sampling and present the higher order formalism. Then we describe the data used in this work and the numerical tests performed on them. Finally we present our summary and conclusions.

\section{Method}
\label{sec:method}

For the sake of completeness we will recap the lognormal-Poisson posterior model within a Bayesian framework, first presented in \citet{argo3}. In particular, we will include a power-law bias description in the equations as introduced in \citet{2015MNRAS.446.4250A}.

\subsection{Bayesian framework}

In a Bayesian inference framework we need to first define the prior of the sought signal $\mbi s$: $\pi(\mbi s)$, and then the likelihood of the data given the signal $\cal L(\mbi d|\mbi s)$. These ingredients permit us to define the posterior distribution function, i.e., the PDF of a signal given the data:
\ba
\label{eq:bayes}
\cal P(\mbi s|\mbi d)\propto\pi(\mbi s)\times\cal L(\mbi d|\mbi s)
\ea

\subsubsection{The Prior}

The signal we want to reconstruct within a Bayesian inference framework is the linear over-density field, thus, $\mbi s\equiv\mbi \delta_{\rm L}$. From now on, we will consider a regular grid with a cubical volume $V$ of side $L$ subdivided into $\rm N_{c}$ cells.
We assume as a prior that $\mbi \delta_{\rm L}$ is Gaussian distributed with zero mean
\ba
\label{prior}
\pi(\mbi\delta_{\rm L}\mid \mat C_{\rm L})=\frac{1}{\sqrt{(2\pi)^{N_{c}}\det(\mat C_{\rm L})}}\exp\left(-\frac{1}{2}\mbi \delta^{\rm T}_{\rm L}\mat C_{\rm L}^{-1}\mbi\delta_{\rm L}\right)\,,
\ea
where $\mat C_{\rm L}=\langle\mbi\delta^{\rm T}_{\rm L}\mbi\delta_{\rm L}\rangle$ is the  co-variance matrix.

\subsubsection{The Likelihood}

The likelihood defines the model of the data. One has to include here the connection between the signal $\mbi s$ and the data $\mbi d$. This is achieved with  a structure formation model for the dark matter field, relating the primordial linear over-density field $\mbi \delta_{\rm L}$ to the cosmic evolved one $\mbi \delta$, and a biasing prescription relating $\mbi \delta$ to the galaxy population. 


\subsubsection*{The dark matter density field}

In this work, we relate the non-linear over-density field $\mbi\delta=\mbi\rho/\bar{\mbi\rho}-1$ (with $\mbi\rho$ being the density) through a logarithmic transformation to linear
\ba
\mbi\delta_{\rm L}=\log(1+\mbi\delta)-\mbi\mu\,,
\ea
where 
\ba
\mbi\mu = \langle \log(1+\mbi\delta)\rangle\,.
\ea
This yields the  lognormal model for the density field.
It is particularly interesting due to its rich cosmological information content \citep[][]{2014MNRAS.439L..11C}. Looking carefully at its derivation from the continuity equation applied to a cosmic fluid, one finds that it is valid for a Lagrangian co-moving framework before shell crossing \citep{1991MNRAS.248....1C,2012MNRAS.425.2443K}.
This implies that the lognormal assumption, applied to cosmic evolved density fields observed in Eulerian coordinates, is not accurate, especially in the three-point statistics \citep{2014MNRAS.437.2594W,2015MNRAS.452..686C}, although it gives a fair description of the two-point statistics \citep[][]{2009ApJ...698L..90N}.
Nevertheless, we will use this prior in this work as a reference to study the efficiency of the sampler, neglecting displacements connecting Lagrangian to Eulerian space, as it would be required for an accurate structure formation description.
We note, however, that more complex structure formation models can be implemented. Sampling the Lagrangian tracers with the displacement field given an arbitrary structure formation model within a Gibbs-sampling framework, the lognormal assumption turns out to become reasonable, and for $|\delta|\ll1$ it ultimately tends towards the Gaussian PDF \citep{2019arXiv191100284K}. 
 The lognormal assumption, however, ensures positive definite densities, i.e. $\rho\geq0$. This is very important, since otherwise one has to cut-off within a general non-linear bias description cells with $\rho<0$ resulting in an artificial lack of power of the density field.


 \subsubsection*{Link between the dark matter field and the galaxy distribution}

It is natural to define the data as the number counts of galaxies $\mbi d\equiv\mbi N^{g}$ in the above defined regular grid, as this allows for a clear statistical description, and for efficient operations relying on fast Fourier transforms. 

To capture the discrete nature of the data we can assume a Poisson likelihood, which was introduced to Bayesian reconstruction in cosmology in \citet{argo,argo3}
\ba
\label{likelihood}
\mathcal{L} (N_{k}^{g}\vert\lambda_{k})=\prod_{k}\frac{(\lambda_{k})^{N_{k}^{g}}e^{-\lambda_{k}}}{N_{k}^{g}!}\,,
\ea
where the expected number counts per cell $k$ is given by
\ba
\lambda_{k}=f_{N}w_{k}(1+\delta_{k})^{b}\,,
\ea
$f_{N}$ is the normalization of the ensuring a given number density $\bar{N}$, $w_k$ is the three-dimensional completeness at cell $k$, and  $b$ is the power-law bias parameter  \citep[see][]{2014MNRAS.439L..21K,2015MNRAS.446.4250A}. 

Prior reconstructions considered only the variance of the Poisson distribution within a Gaussian likelihood, which does not ensure positive definite density fields in the reconstruction \citep[][]{1995ApJ...449..446Z}.
We note that this is the simplest discrete PDF we can consider without requiring any additional parameter. 
In general, the distribution of galaxies is not Poisson distributed
 \citep[see][]{1980lssu.book.....P}.
 There are some PDFs which can capture the deviation from Poissonity \citep[see e.g.][]{1989ApJ...341..588S,1998MNRAS.299..207S,2014MNRAS.439L..21K,2014MNRAS.441..646N,2015MNRAS.450.1486A}, that can be implemented in a Bayesian framework \citep{2015MNRAS.446.4250A}.
 
\subsubsection{The Posterior}

Based on the prior and likelihood defined in  the previous sections we can now define the posterior PDF (see equation   \ref{eq:bayes}).
For convenience, let us write the negative logarithm of the posterior as
\ba
\label{eq:posterior}
-\ln \mathcal{P} =-\ln \pi-\ln \mathcal{L}.
\ea
This permits us to write the prior term \ref{prior}  as
\ba
\label{lnprior}
-\ln \pi(\delta_{\rm L}\mid \mat C_{\rm L})=\frac{1}{2}\mbi \delta^{\rm T}_{\rm L}\mat C_{\rm L}^{-1}\mbi \delta_{\rm L}+c\,,
\ea
where we have included terms that do not depend on the signal in the term $c$.
The negative logarithm of the likelihood, taking equation   \ref{likelihood}, is simply
\ba
\label{lnlikelihood}
-\ln \mathcal{L}(N_{k}^{g}\vert\lambda_{k})=\sum_{k}\lambda_{k}-N_{k}\ln\lambda_{k}+c'\,,
\ea
with $c'\neq c'(\mbi\delta_{\rm L})$.
This permits us to compute the gradients with respect to the signal of the prior and the likelihood in a straightforward way as introduced in \citet{argo3}.
For the prior we obtain
\ba
\label{eq:grad1}
-\frac{\partial \ln \pi}{\partial \mbi\delta_{\rm L}}=\mat C_{\rm L}^{-1} \mbi\delta_{\rm L}\,.
\ea
And for the likelihood we use the chain rule
to get
\ba
\label{eq:grad2}
-\frac{\partial \ln \mathcal{L}}{\partial \delta_{L,i}}=b\left(\lambda_{i}-N_{i}\right)\,.
\ea
These gradients permit us to compute  either the maximum a posteriori, when solving the corresponding equation set to zero, or to sample from the posterior PDF using the Hamiltonian Monte Carlo sampling, as we will show in the next section.

\subsection{Hamiltonian Monte Carlo sampling}
\label{sec:HMC}
To sample the posterior we rely on the Hamiltonian Monte Carlo sampling technique (HMC) \citep{DUANE1987216,Neal,fasthamiltonian}. Let us recap the method in this section and extend it to higher orders. 
The Hamiltonian is defined as a function of the generalized phase space coordinates of positions  $\mbi q$ and  momenta $\mbi p$, through the potential energy $\mathcal{U}(\mbi{q})$ and the kinetic energy $\mathcal{K}(\mbi{p})$
\ba
\label{eq:hamil}
\mathcal{H}(\mbi{q},\mbi{p})=\mathcal{U}(\mbi{q})+\mathcal{K}(\mbi{p})\,.
\ea
The kinetic energy is expressed as
\ba
\label{eq:mom}
\mathcal{K}(\mbi{p})=\frac{1}{2}\mbi{p}^{\rm T}\mat{M}^{-1}\mbi{p}\,,
\ea
where $\mat{M}$ is the mass matrix, describing the co-variance of the momenta. It represents the degree of freedom in the Hamiltonian sampler, and its structure can be crucial for the efficiency \citep{MCMC}.
One chooses an adequate mass matrix, encoding both the prior and the likelihood information. In general, such a mass matrix will be non-diagonal, and there are ways of implementing them in an efficient way \citep{2019arXiv191100284K}.
In this work we will restrict our studies to a full volume, for which a mass matrix given by the inverse matter co-variance matrix is nearly optimal, $\mat M=\mat C_{\rm L}^{-1}$ \citep[see][]{2008MNRAS.389.1284T}.


To relate the Hamiltonian dynamics to a probabilistic measure, we resort to the canonical distribution definition
\ba
\label{eq:canonical}
\mathcal{P}(\mbi{q},\mbi{p})=\frac{1}{Z}\,{\rm e}^{-\mathcal{H}(\mbi{q},\mbi{p})}\,,
\ea
where $Z$ is the normalization of the distribution function. 
The latter equation can also be expressed as
\ba
\mathcal{P}(\mbi{q},\mbi{p})=\mathcal{P}(\mbi{q})\mathcal{P}(\mbi{p})=\frac{1}{Z}\,{\rm e}^{-\mathcal{U}(\mbi{q})}\,{\rm e}^{-\mathcal{K}(\mbi{p})}\,,
\ea
according to our previous definitions, factorized into two separated probabilities corresponding to the potential energy (and the positions): $\mathcal{P}(\mbi{q})$, and to the kinetic energy (and the momenta): $\mathcal{P}(\mbi{p})$. 
It is interesting now to identify the potential energy $U(\mbi{q})$ with the negative logarithm of the posterior distribution function (equation   \ref{eq:posterior})
\ba
\label{eq:pos}
\mathcal{U}(\mbi{q})=-\ln \mathcal{P}(\mbi{q})\,,
\ea
and realise that the kinetic term $K(\mbi{p})$ defines a multivariate Gaussian distribution function
\ba
\mathcal{P}(\mbi{p})\propto e^{-\mathcal{K}}=e^{-\frac{1}{2}\mbi{p}^{\rm T} \mat M^{-1}\mbi{p}}\,.
\ea
This implies that the Hamiltonian Monte Carlo sampling only requires a Gaussian field with a free Hamiltonian mass to sample arbitrary non-Gaussian PDFs.
We can now further identify  the positions, $\mbi q$, as the variable to sample, i.e., the sought signal, in our case, the primordial fluctuations $\mbi \delta_{\rm L}$. The momenta, $\mbi  p$, are artificially introduced in the kinetic term just to allow us to explore the phase-space, therefore, to evolve the system and get $\mbi q$. The marginalization is done to avoid the dependence on the momenta when obtaining the posterior. This is achieved by randomly drawing new momenta in  each iteration, disregarding the ones of the previous step.

The partial derivatives of the Hamiltonian determine how $\mbi{q}$ and $\mbi{p}$ change with time, $t$, according to Hamilton's equations of motion 
\ba
\label{eq:ham1}
\frac{\rm d \mbi q}{\rm d t} & = & \frac{\partial {\mathcal H}(\mbi{q},\mbi{p})}{\partial \mbi p}=-\{\mathcal{H},\mbi q \}\,,\\
\label{eq:ham2}
\frac{\rm d \mbi p}{\rm d t} & = & -\frac{\partial {\mathcal H}(\mbi{q},\mbi{p})}{\partial \mbi q}=-\{\mathcal{H},\mbi p \}\,,
\ea
where we have introduced the Poisson bracket definition: $\{f,g\}=\left(\frac{\partial f}{\partial \mbi q}\right)^{\rm T}\frac{\partial g}{\partial \mbi  p}-\left(\frac{\partial f}{\partial \mbi p}\right)^{\rm T}\frac{\partial g}{\partial \mbi  q}$,  for reasons which will be clear below.
Taking into account equations    \ref{eq:hamil} and \ref{eq:mom}, Hamilton's equations can then be written as
\ba
\label{eq:gradp}
\frac{\rm d\mbi q}{\rm d t}  & = & \mat{M}^{-1}\mbi p \,,\\
\label{eq:gradu}
\frac{\rm d\mbi p}{\rm d t} & = &-\frac{\partial \mathcal{U}(\mbi{q})}{\partial \mbi q}\,.
\ea
The particular expression to the latter equation, in our case study, is given by the sum of equations    \ref{eq:grad1} and \ref{eq:grad2}.

Moreover, the Hamiltonian dynamics has to fulfill a series of properties:
\begin{itemize}
\item The Hamiltonian $\mathcal{H}$ is conserved as $\mbi q$ and $\mbi p$ evolve through time: $\frac{\partial \mathcal{H}}{\partial t}=0$.
\item The dynamics also preserves the phase space volume according to Liouville's theorem.
\item Hamiltonian dynamics is reversible, i.e., mapping from a state to the next state is bijective  one-to-one, and therefore, the inverse mapping is obtained by changing the sign in the time derivatives in equations    \ref{eq:ham1} and \ref{eq:ham2}.
\end{itemize}
These properties  together imply that the canonical distribution is invariant with respect to any transformation.

{\color{black}
\subsubsection{Discretisation and efficiency}
\label{sec:second1}

To evolve the Hamiltonian system numerically, we must discretise the Hamilton's equations of motion using a finite time step, and thus introducing an inevitable error.
An explicitly time-invariant Hamilton function is energy conserving. Thus, the Hamiltonian difference  $\Delta \mathcal{H}(\mbi q,\mbi p)=\mathcal{H}(\mbi q',\mbi p')-\mathcal{H}(\mbi q,\mbi p)$ between the old $(\mbi q, \mbi p)$ and new $(\mbi q', \mbi p')$ proposed state of the system should vanish. Nonetheless, numerical errors in the discretisation scheme violate this conservation.

Due to this numerical error, one has to introduce a Metropolis-Hastings rejection step. The proposed new state obtained by 
\ba
{\cal P}_{\rm acceptance}=\min\left[1,{\rm e}^{-\Delta \mathcal{H}(\mbi q,\mbi p)}\right]\,,
\ea
where  $\Delta \mathcal{H}(\mbi q,\mbi p)=\mathcal{H}(\mbi q',\mbi p')-\mathcal{H}(\mbi q,\mbi p)$ stands for the  difference in the Hamiltonian between the old $(\mbi q, \mbi p)$ and new $(\mbi q', \mbi p')$ proposed state of the system.

The chosen time-step and discretisation scheme will have a great impact on the acceptance rate and the computational efficiency.

Since the energy is an extensive quantity, the total error will grow with an increasing number of dimensions \citep[see][]{2010arXiv1001.4460B,2017arXiv171105337B}. For this reason, for a given error per dimension, there is a number of dimensions from which on higher order schemes become more efficient than the standard second-order discretisation scheme allowing for larger step-sizes with higher acceptance rates.  We will therefore investigate our physical problem at different resolutions in section \ref{sec:numerical}. 
Let us revise these schemes first from a theoretical perspective.

}

\subsubsection{Second-order discretisation}
\label{sec:second}

Let us follow the formalism of   \citet{1988PhRvD..38.1228C,creutz,CAMPOSTRINI1990753}.  
We start with the  basic leap-frog algorithm.
For a Hamiltonian of the generalized coordinates $\mbi q, \mbi p$: 
$\mathcal{H}(\mbi q, \mbi p)$, we can define the translation operator $\mathcal{T}(\epsilon)$, evolving the system along a time step of size $\epsilon$. Due to the property listed in the previous subsection on  reversibility of the Hamiltonian, we demand
\ba
\label{eq:timerev}
\mathcal{T}^{-1}(\epsilon) = \mathcal{T}(-\epsilon) \, .
\ea
We can now split the time translation into separated parts acting on $\mbi p$ and $\mbi q$ individually 
\ba
\mathcal{T}_q(\epsilon): (\mbi q, \mbi p) &\rightarrow& (\mbi q', \mbi p)\, , \\
\mathcal{T}_p(\epsilon): (\mbi q, \mbi p) &\rightarrow& (\mbi q, \mbi p')\, , 
\ea
where the new states $\mbi q'$ and $\mbi p'$ are given according to the equations of motion \ref{eq:gradp} and \ref{eq:gradu} with
\ba
\mbi q' &=& \mbi q +\epsilon\,\mat M^{-1}\mbi p  \, , \\
\mbi p' &=& \mbi p - \epsilon\,\frac{\partial \mathcal{U}}{\partial \mbi q}  \, .
\ea
Following equations    \ref{eq:ham1} and \ref{eq:ham2} the evolution of system from an old state $(\mbi q, \mbi p)$ to a new one $(\mbi q', \mbi p')$ is obtained through the action of the Hamilton operator on $(\mbi q, \mbi p)$.
A naive translation of step-size $\Delta\tau=\epsilon$, such as $\mathcal{T}\,(\epsilon) = \mathcal{T}_p(\epsilon) \mathcal{T}_q\,(\epsilon)$, will violate time reversibility since
\ba
( \mathcal{T}_p(\epsilon) \mathcal{T}_q(\epsilon))^{-1} =\mathcal{T}_q(\epsilon)^{-1} \mathcal{T}_p(\epsilon)^{-1} \neq   \mathcal{T}_p(\epsilon)^{-1} \mathcal{T}_q(\epsilon)^{-1} \, .
\ea
An obvious choice of a time translation operator to preserve reversibility can be constructed by symmetrizing the operator
\ba
\label{eq:2ndorder}
\mathcal{T}_2(\epsilon) = \mathcal{T}_p(\epsilon/2)\mathcal{T}_q(\epsilon) \mathcal{T}_p(\epsilon/2) \, ,
\ea
which is the commonly used  leap-frog discretisation scheme. It preserves phase space volume, and is also time reversible. {\color{black} We favour this formulation with respect to $\mathcal{T}_q(\epsilon/2)\mathcal{T}_p(\epsilon) \mathcal{T}_q(\epsilon/2)$ (interchanging the role of $p$ with $q$). It permits us to  save some matrix inversions when computing $\mat M^{-1}\mbi p$\footnote{\color{black} Within the \texttt{COSMIC BIRTH} code this operation involves more operations (including convolutions) as explained in the appendix in \citet{2019arXiv191100284K}.}. This becomes clear further below (whether, despite of these extra computations, this formulation is more efficient needs to be investigated in future work).}
A single iteration calculates approximations to the position and momenta at time $t+\epsilon$ from these quantities at $t$ as it follows
\ba
\mbi p\left(t+\frac{\epsilon}{2}\right)&=&\mbi p(t)-\frac{\epsilon}{2}\frac{\partial\mathcal{U}}{\partial\mbi q}(\mbi q(t))\label{eq:firstgrad}\\
\mbi q\left(t +\epsilon\right)&=&\mbi q(t)+\epsilon \,\mat M^{-1} \,\mbi p\left(t +\frac{\epsilon}{2}\right)\\
\mbi p\left(t +\epsilon\right)&=&\mbi p\left(t +\frac{\epsilon}{2}\right)-\frac{\epsilon}{2}\frac{\partial \mathcal{U}}{\partial \mbi q}(\mbi q\left(t +\epsilon\right))\label{eq:lastgrad}\,,
\ea
{\color{black} where the last gradient computation (Eq.~\ref{eq:lastgrad}) serves as the first one (Eq.~\ref{eq:firstgrad}) for the next iteration.  Note that the momenta are replaced after each completed Hamiltonian MC iteration, and therefore the last matrix inversion applied to the momenta cannot always be used for the next iteration in a scheme of the form $\mathcal{T}_q(\epsilon/2)\mathcal{T}_p(\epsilon) \mathcal{T}_q(\epsilon/2)$.}
Eqs.~\ref{eq:firstgrad} to \ref{eq:lastgrad}   correspond to a second-order discretisation of the equations of motion, (${\cal O}(\epsilon^{2})$), as we will discuss in the next section.
  Eqs.~\ref{eq:firstgrad}-\ref{eq:lastgrad}.
In practice, this scheme is applied {\color{black} 
\ba
\label{eq:neval}
N_{\rm eval}^{\rm 2nd}=1+{\rm floor}(u_N\times N_{\rm  eval})
\ea
times with a random time step $\epsilon^{\rm eff}=u_\epsilon\times \epsilon$}, where $u_N$ and $u_\epsilon$ are random numbers, which help the HMC sampler to explore the parameter space {\color{black} avoiding resonant trajectories} \citep[][]{Neal}. 
{ \color{black}
 The corresponding second-order leap-frog scheme is thus given by multiple evaluations of the equations of motion
\ba
N_{\rm eval}^{\rm 2nd}\times\left(\mathcal{T}_2(\epsilon^{\rm eff}) = \mathcal{T}_p(\epsilon^{\rm eff}/2)\mathcal{T}_q(\epsilon^{\rm eff}) \mathcal{T}_p(\epsilon^{\rm eff}/2) \right)\,.
\ea
Hence, the global  step-size is given by 
\ba
\label{eq:Tau2ndorder}
\Delta\tau_{\rm 2nd}=N_{\rm eval}^{\rm 2nd}\times \epsilon^{\rm eff}=(1+{\rm floor}(u_N\times N_{\rm  eval}))\times u_\epsilon\times \epsilon\,.
\ea
}

\subsubsection{Higher order discretisation}

\label{sec:higher}

Let us now revise Hamilton mechanics to find a generalization of the leap-frog integration beyond second-order.  
For any conserved function, which depends on the phase-space variables at time $t$, $f(t, \mbi q, \mbi p)$, the time derivative vanishes
\ba
\frac{\mathrm d f} {\mathrm d t} =0\,,
\ea
and hence
\ba 
\frac{\partial f}{\partial t}+\frac{\partial f}{\partial \mbi q}\frac{\mathrm d \mbi q}{\mathrm d  t}+\frac{\partial f}{\partial \mbi p}\frac{\mathrm d \mbi p}{\mathrm d   t}=\frac{\partial f}{\partial t} -\left\{\mathcal{H},f\right\} &=& 0\,,
\ea
where we have used equations    \ref{eq:ham1}, \ref{eq:ham2}, and the Poisson bracket definition.
Thus we can identify the partial time derivative to the Hamilton operator 
\ba 
\label{eq:der}
\frac{\partial}{\partial t} \,f &=& \{\mathcal{H}, \cdot  \} \, f  \,.
\ea
From this we can write a time evolution of $f$ from time $t$ to $t+\epsilon$ by following transformation
\ba
\label{eq:expf}
f(t+\epsilon, \mbi q, \mbi p) = \mathrm{e}^{\mathcal{H}\epsilon} f(t, \mbi q, \mbi p)\, ,
\ea
which is the classical equivalent to the time evolution solution for the Schr{\"o}dinger equation in quantum mechanics.
Expanding $f(t, \mbi q, \mbi p)$ as a function of the time evolution in a Taylor series, we can write
\ba
f(t+\epsilon, \mbi q, \mbi p) =& f + \frac{\partial  }{\partial t} \,f\,\epsilon + \frac{1}{2} \frac{\partial^2 }{\partial t^2} \,f\,\epsilon^2 + \frac{1}{6} \frac{\partial^3 }{\partial t^3} \,f\, \epsilon^3 +\dots  \, ,\nonumber \\
\label{eq:expansionf}
=& f + \{\mathcal{H},\cdot\} \,f\,\epsilon +\frac{1}{2} \{\mathcal{H},\{\mathcal{H},\cdot\} \} \,f\, \epsilon^2+\nonumber \\ & \frac{1}{6}   \{\mathcal{H},\{\mathcal{H},\{\mathcal{H},\cdot\} \} \}  \,f\,\epsilon^3 +\dots \, ,
\ea
where we have used  equation   \ref{eq:der}.
We can relate equation   \ref{eq:expf} to the series expansion given by equation   \ref{eq:expansionf} and write 
\ba
\label{eq:hrel}
\mathrm{e}^{\mathcal{H}\epsilon} = \mathcal{T}_n(\epsilon) - \Delta_{n+1}\epsilon^{n+1}+ \mathcal{O}(\epsilon^{n+2})\,,
\ea
where the errors of the time evolution operator $\mathcal{T}_n$, with respect to the analytic solution, $-\Delta_{n+1}\epsilon^{n+1}$, are of order ${n+1}$. 
The operator $\mathcal{T}_n$ is made out of a number of concatenated  Hamiltonian operators depending on the order $n$. This suggests the idea to construct higher order Hamiltonian schemes based on recursive applications of lower order ones.
A naive fourth-order scheme could  then be constructed as the application of two successive second-order ones. Let us define with that spirit our tentative $(n+2)$-order operator as
\ba
\label{eq:tent}
\mathcal{T}_{n+2}(2i\epsilon)=\mathcal{T}_{n}^{\ i}(\epsilon)\mathcal{T}_{n}^{\ i}(\epsilon)\,,
\ea
which based on equation   \ref{eq:hrel} can be expanded to
\ba
\lefteqn{\mathcal{T}_{n+2}(2i\epsilon)=\mathrm{e}^{\mathcal{H}2i\epsilon}+\Delta_{n+1}2i\epsilon^{n+1}+}\\&&\Delta_{n+2}\epsilon^{n+2}+\mathcal{O}(\epsilon^{n+3})\nonumber\,,
\ea
where $i$ is the number of times the operator $\mathcal{T}_{n}$ is successively applied, and $\Delta_{m}$  stands for the error factors at different order $m=1,2,\dots$. We have kept only the first term of the Taylor expansion of $\mathrm{e}^{\mathcal{H}i\epsilon}$,  whenever it appeared  multiplying error terms to correctly keep track of the orders. This implies that the naive successive concatenation of second-order leap-frog operations does not yield a fourth-order accurate scheme.
The problem here is the presence of error terms of order below  $\mathcal{O}(\epsilon^{n+3})$. 
Let us focus first on the $(n+1)$-order term  $\Delta_{n+1}2i\epsilon^{n+1}$. 
The solution proposed by \citet{creutz} and \citet{CAMPOSTRINI1990753} consists of introducing a backward step to exactly cancel out the $(n+1)$-order error term, which necessarily needs to have a step-size of
\ba
-s\epsilon = -(2i)^{1/(n+1)}\epsilon\,.
\ea
From  equation   \ref{eq:hrel} we can verify that an opposite error term $-\Delta_{n+1}2i\epsilon^{n+1}$ is obtained.
To see how the $(n+2)$-order term vanishes we need to construct a time-reversible operator (see equation   \ref{eq:timerev}), for which 
$\mathcal{T}(-\epsilon) \mathcal{T}(\epsilon)=1$ holds within the order of the scheme.
Inserting the expansions from equation   \ref{eq:hrel} with the ansatz of equation   \ref{eq:tent} we get
\ba
\lefteqn{\mathcal{T}_{n+2}(-\epsilon)\mathcal{T}_{n+2}(\epsilon)=1+}\\&&\Delta_{n+1}(\epsilon^{n+1}-\epsilon^{n+1})+\Delta_{n+2}(\epsilon^{n+2}-\epsilon^{n+2})+\mathcal{O}(\epsilon^{n+3})\nonumber\,.
\ea
Focusing now on the $(n+2)$-order term we find that it can vanish for odd numbers of $n$, since then  $\epsilon^{n+2}-\epsilon^{n+2}$ cancels out, however, with the term $\Delta_{n+2}\epsilon^{n+2}$ not having to be zero. Only for even numbers of $n$, and in order to accomplish the reversibility condition, we can state that $\Delta_{n+2}$ vanishes, and hence also the $(n+2)$-order error term.
 For this reason, \citet{creutz} suggested the following recursive scheme for even numbers of $n$
\ba
\label{eq:highorder}
\mathcal{T}_{n+2}((2i-s)\epsilon)=\mathcal{T}^{\ i}_{n}(\epsilon)\mathcal{T}_{n}(-s\epsilon)\mathcal{T}^{\ i}_{n}(\epsilon)\,,
\ea
where reversibility and phase-space volume conservation are accomplished. 
Hence, iterating this scheme recursively produces a discretisation of Hamilton's equations of motion to any desired even order. 
{\color{black} To make a fourth-order scheme one needs $2i+1$ times computations of the basic second-order leap-frog scheme. The corresponding global step-size is given by
\ba
\label{eq:Tau4thorder}
\Delta\tau_{\rm 4th}=(2i-s(n=2))\,\epsilon^{\rm eff}=(2i-(2i)^{1/(3)})\,\epsilon^{\rm eff}\,,
\ea
with a random time step $\epsilon^{\rm eff}=u_\epsilon\times \epsilon$, as in the second-order case.
It is important to stress that, applying the second-order leap-frog as usual, making large numbers of evaluations ($N_{\rm eval}^{\rm eff}$) of Hamiltonian equations of motion within the fourth-order method (equation   \ref{eq:highorder}), will result in an inefficient scheme. Instead, we explore in this work the application of this recursive formula calling the second-order discretisation only  $2i+1$ times, and we explore the optimal $i$ number. The aim of this is to explore the possibility of making larger time steps at a comparable computational cost, and hence gain efficiency sampling from the posterior PDF.
}

\begin{table*}
\centering
\begin{tabular}{|c|l|l|l|l|l|l|}
\hline \hline
\multicolumn{1}{|l|}{Use} & Host   & CPU                        & \multicolumn{1}{c|}{Freq} & Cores & RAM    & Disk  \\ \hline \hline
Login \& GPU node         & deimos & 2x Intel Xenon E5-2630 v4  & 2.20 GHz                  & 20    & 1TB    & 11 TB \\ \hline
Computing node            & diva   & 12x Intel Xenon E5-2630 v4 & 2.10 GHz                  & 192   & 4.5 TB & 40 TB \\ \hline
\end{tabular}
\caption{Deimos/Diva characteristics.}
\label{deimos}
\end{table*} 

\begin{figure*}
\centering
\begin{tabular}{cc}
\includegraphics[width=8cm]{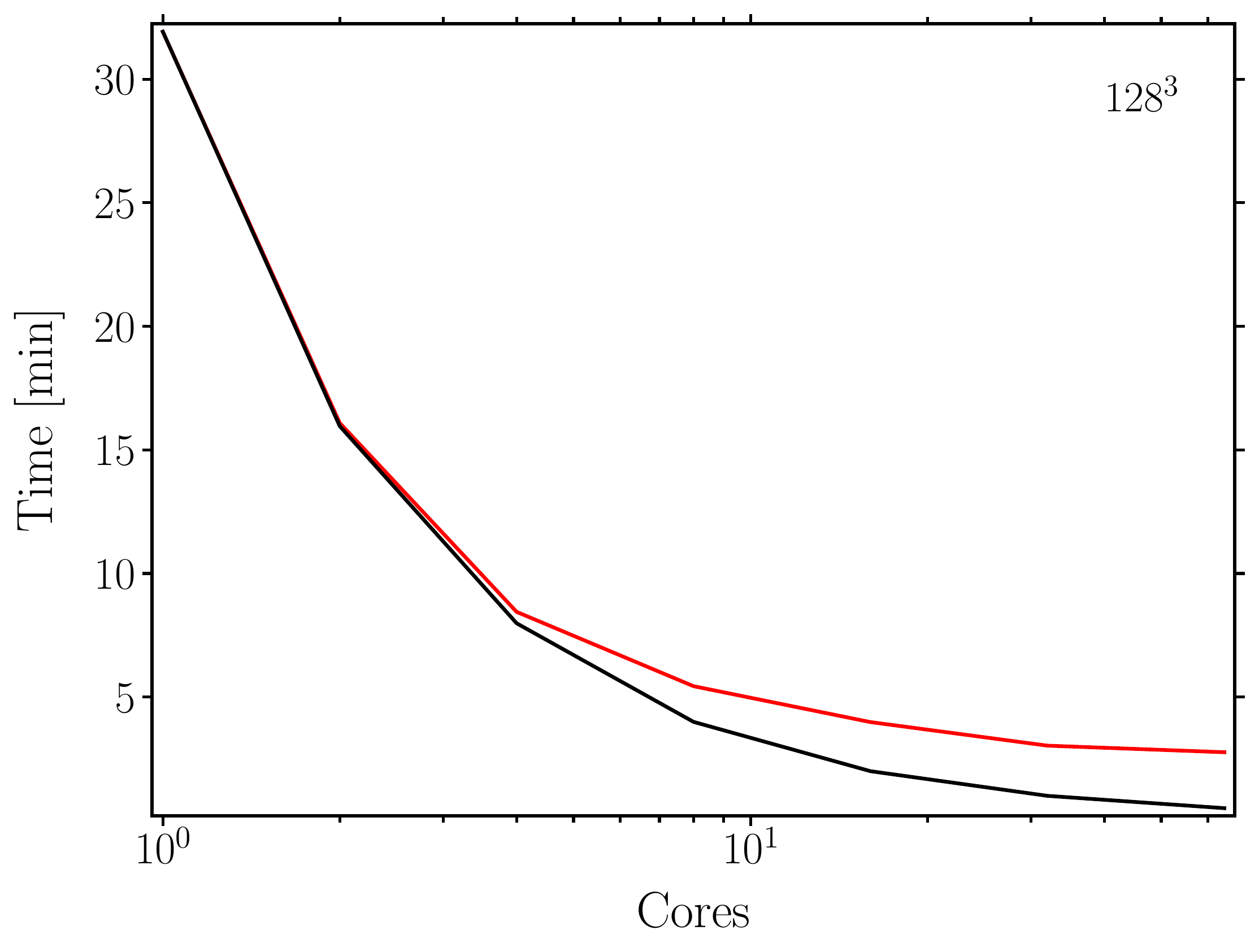}
\hspace{0.5cm}
\includegraphics[width=8.1cm]{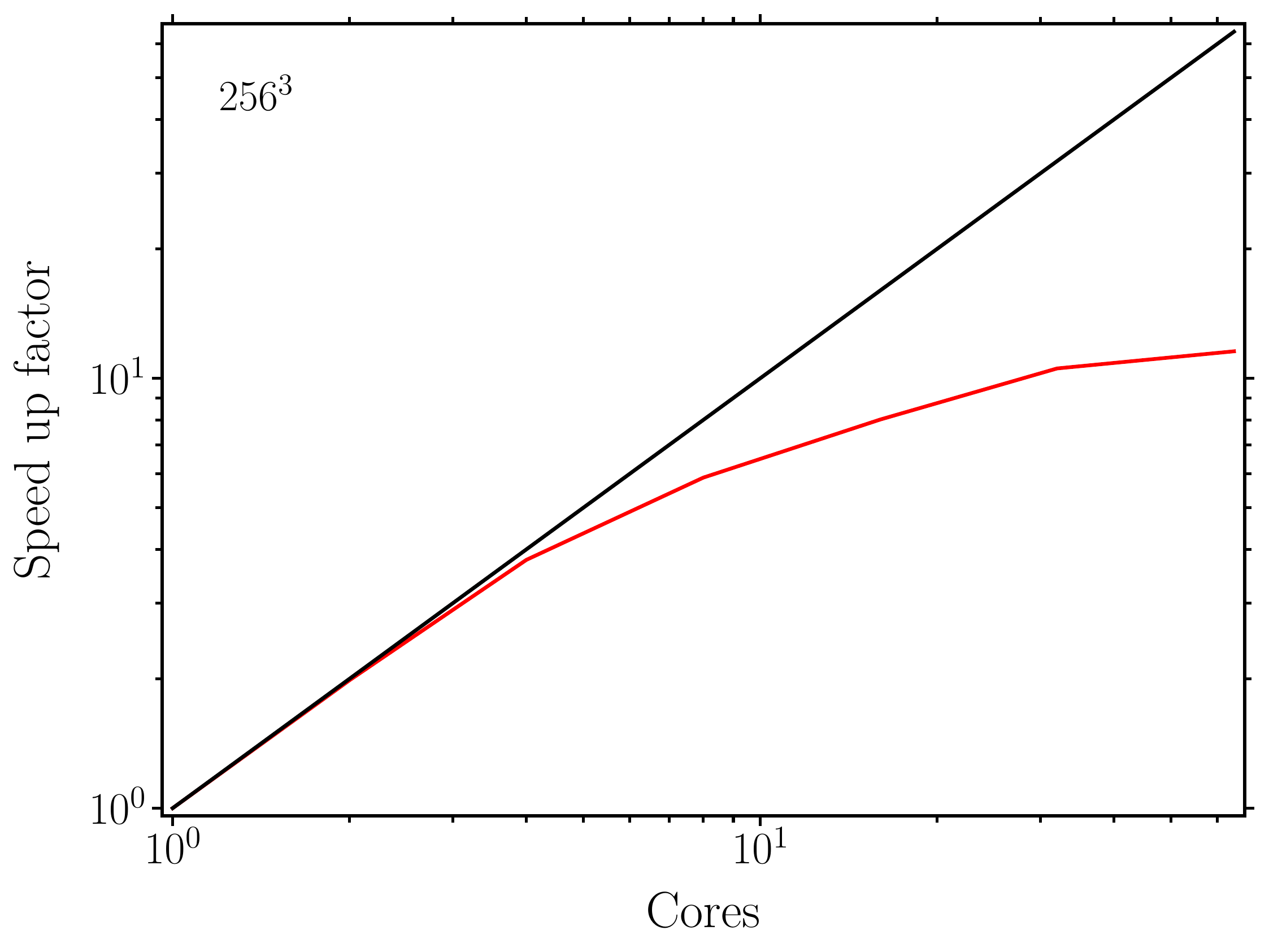}
\end{tabular}
\caption{\label{fig:cores} {\bf The panel on the left} (used for the 128$^3$ case) shows the computation time as a function of the number of cores (solid red line). The solid black line represents the reference ideal case, in which the computation time decreases to the half each time we double the number of cores. {\bf The panel on the right} (used for the 256$^3$ case) shows the speed up factor as a function of the number of cores. The solid black line represents the ideal case, in which the speed up factor is $1$ for $1$ core, $2$ for $2$ cores and so on.}
\end{figure*}

\section{Numerical validation}
\label{sec:numerical}

In this section we show our parameter study, analyzing the optimal setting for the fourth-order leap-frog algortihm, as compared to the second-order discretisation scheme.
We start exploring the parameter space on a lower resolution, and then we focus on a number of constrained configurations on a set of higher resolution runs.
Based on this we will make a robust assessment of the convergence of the chains and the corresponding correlation lengths.

\subsection{Data used in this work}
\label{sec:mocks}

To validate the method we restrict this analysis to a mock galaxy catalog corresponding to a single snapshot at $z=0.57$. In particular, it matches  the CMASS sample of luminous red galaxies (LRGs), which is a complete sample, nearly constant in mass and volume, limited between the redshifts $0.43\le z \le 0.7$ (see \citet{Andersonboss} for details of the targeting strategy). We use the $N$-body based mock galaxy catalog constructed to match the clustering bias and number densities of the BOSS DR12 CMASS galaxies at the mean redshift of $\bar{z}=0.57$.


The mock galaxy catalog used in this study was presented in \citet[][]{2016MNRAS.460.1173R} and was  extracted from the BigMDPL N-body simulation\footnote{See https://www.cosmosim.org/cms/simulations/bigmdpl/}, one of the Multidark simulation project, which was performed using the \texttt{GADGET-2} code \citep{2005MNRAS.364.1105S}. The BigMDPL was run with $3.840^3$ particles on a volume of $(2.5\,h^{-1}{\rm Gpc}$ $)^3$ assuming $\Lambda$CDM Planck cosmology with \{$\Omega_\Lambda =0.6928, \Omega_{\rm M}=0.307, \Omega_{b}=0.0482,\sigma_8=0.828,n_s=0.961$\}, and a Hubble constant ($H_0=100\,h\,\kmsmpc$) given by  $h=0.677$.  Halos and subhalos were identified  using the \texttt{ROCKSTAR} halo finder \citep[][]{2013ApJ...762..109B}.

\subsection{Results}

\label{sec:study}

For our study we rely on the \texttt{COSMIC BIRTH} code \citep{2019arXiv191100284K} to sample the density field with the lognormal-Poisson model, switching off:  displacements, peculiar motions, and selection effects. This corresponds to the first Gibbs-sampling step listed in the \texttt{COSMIC BIRTH} paper, which initially represented the bottle-neck of the computations.  The posterior distribution function is sampled with the HMC sampling technique following the methods described in the previous section, including an automatic estimation of the logarithmic mean field $\mu$ \citep[see][]{KitauraGalleraniFerrara2012}.



We choose first two  meshes of $128^3$ and  of $256^3$ on a cubical volume of 1250 $h^{-1}$ Mpc side. We perform  a nearest-grid-point mass assignment of the mock galaxy catalog on the grid to obtain the data array, as the number counts per cell.

{\color{black}At the final part of our analysis, we consider also meshes of 64$^3$ to verify how second-order schemes start becoming more efficient towards lower statistical dimensions.}

\subsubsection{Parallelisation and optimal number of cores}

The numerical tests have been performed using the \textit{Diva Severo Ochoa} machine, which is a High Performance Computer at the IAC with specifications shown in table \ref{deimos}. 

First, a study of the optimal number of cores to run the Open-MP parallel \texttt{COSMIC BIRTH} code is presented. To do so, the code has been run for the same parameters ($i$, step-size $\epsilon$, seed and number of iterations) for different number of cores: $1,2,4,8,16,32$, and $64$.

\begin{itemize}
 
 \item Low resolution case: 128$^3$ cells.

The left panel in figure \ref{fig:cores} shows the computational time needed to reach $100$ iterations as a function of the number of cores, represented by the the red line. The black line is the reference one from a perfect scaling of the computation time with the number of cores, which means that the computational time decreases to the half each time we double the number of cores.
As we can see, for more than $8$ cores, the computation time decreases slowly until it becomes almost constant for more than $32$ cores. Hence, the computation time saved using $16, 32$ or $64$ cores is not remarkable enough compared to using $8$, as it deviates from the ideal case (black line). For this reason all the runs with $128^{3}$ cells in this study were performed with 8 cores.
\\
 \item High resolution case: 256$^3$ cells.
The right panel in figure \ref{fig:cores} represents the speed up factor by the solid red line, which is defined by the largest time of all runs (the one for $1$ core) divided by the time of each run. The solid black line shows the reference curve for an ideal speed up factor: $1$ for $1$ core, $2$ for $2$ cores, and so on.
In this case, until $32$ cores, we find that the speed up factor goes approximately as the ideal case. However, for $64$ cores we can see that there is a deviation with respect to the solid black line. For this reason, we choose $32$ cores for the high-resolution in this study. We have chosen a different  representation here as for the low resolution case, to better assess the saturation for large number of cores.
\end{itemize}

{\color{black} We note, however, that our study is not affected by the chosen architecture, as we will express the efficiency as a function of the number of evaluations of the Hamiltonian equations of motion, which is directly related to the number of gradient computations (see \S \ref{sec:HMC}).}

\subsubsection{Convergence criteria}
\label{sec:conv}

\begin{table}
\centering
\begin{tabular}{|c|c|c|c|}
\hline \hline
\begin{tabular}[c]{@{}c@{}}step-size\\ $m\epsilon$ \end{tabular}     & \begin{tabular}[c]{@{}c@{}}Iteration of \\ convergence\end{tabular} & \begin{tabular}[c]{@{}c@{}} CONV \\ $[$NoE$]$\end{tabular}&  Acceptance \\
\hline \hline
\multicolumn{4}{|c|}{$i=1$}                \\ \hline
{\color{black} $0.5\epsilon$ }   & {\color{black} $2530$  }                                                                       & {\color{black}  $7722$  }                                                                & {\color{black}  $98.27\%$ }    
\\ \hline
$\epsilon$   & $650$                                                                        & $2012 $                                                                 & $97.66\%$ 
\\ \hline
$2 \epsilon$  & $250$                                                                        & $916 $                                                                 & $83.51\%$    
\\ \hline
$4 \epsilon$  & $230$                                                                        & $1544 $                                                                 & $51.03\%$    
\\ \hline
$6 \epsilon$  & $250$                                                                        & $2460 $                                                                 & $33.00\%$    
\\ \hline
$8 \epsilon$  & $258$                                                                        & $3605 $                                                                 & $24.05\%$    
\\ \hline
$10 \epsilon$ & $232$                                                                       & $4003 $                                                                 & $19.09\%$
\\ \hline
\multicolumn{4}{|c|}{$i=2$}
\\ \hline
{\color{black} $0.5\epsilon$ }  & {\color{black} $245$  }                                                                      & {\color{black} $1246$ }                                                                 & {\color{black} $98.46\%$ } \\ \hline
$\epsilon$   & $68$                                                                        & $366 $                                                                 & $93.30\%$ 
\\ \hline
$2 \epsilon$  & $53$                                                                        & $533$                                                                 & $52.47\%$    
\\ \hline
$4 \epsilon$  & $42$                                                                        & $740$                                                                 & $29.92\%$    
\\ \hline
$6 \epsilon$  & $46$                                                                        & $1136 $                                                                 & $19.92\%$    
\\ \hline
$8 \epsilon$  & $46$                                                                        & $1660 $                                                                 & $16.93\%$    
\\ \hline
$10 \epsilon$ & $36$                                                                       & $1892 $                                                                 & $9.08\%$
\\ \hline
\multicolumn{4}{|c|}{$i=3$}     
\\ \hline
{\color{black} $0.5\epsilon$  } & {\color{black} $209$ }                                                                       & {\color{black} $1485$   }                                                              & {\color{black} $98.13\%$ }
\\ \hline
$\epsilon$   & $32$                                                                        & $360$                                                                 & $76.58\%$ 
\\ \hline
$2 \epsilon$  & $29$                                                                        & $582 $                                                                 & $43.42\%$    
\\ \hline
$4 \epsilon$  & $20$                                                                        & $638 $                                                                 & $23.01\%$    
\\ \hline
$6 \epsilon$  & $24$                                                                        & $1196 $                                                                 & $13.54\%$    
\\ \hline
$8 \epsilon$  & $25$                                                                        & $1672 $                                                                 & $13.96\%$    
\\ \hline
$10 \epsilon$ & $27$                                                                       & $1470 $                                                                 & $11.70\%$
\\ \hline
\multicolumn{4}{|c|}{$i=4$}     
\\ \hline
{\color{black} $0.5\epsilon$ }  & {\color{black}  $100$  }                                                                      & {\color{black} $929$ }                                                                 & {\color{black}  $96.53\%$ } 
\\ \hline
$\epsilon$   & $39$                                                                        & $486 $                                                                 & $81.76\%$ 
\\ \hline
$2 \epsilon$  & $32$                                                                        & $785 $                                                                 & $45.60\%$    
\\ \hline
$4 \epsilon$  & $27$                                                                        & $981 $                                                                 & $26.22\%$    
\\ \hline
$6 \epsilon$  & $26$                                                                        & $1634 $                                                                 & $15.04\%$    
\\ \hline
$8 \epsilon$  & $27$                                                                        & $2342 $                                                                 & $13.66\%$    
\\ \hline
$10 \epsilon$ & $26$                                                                       & $2268 $                                                                 & $6.66\%$
\\ \hline
\end{tabular}
\caption{{\color{black} Convergence as a function of the step-size value, iteration, number of evaluations (NoE) of the Hamiltonian equations of motion, and acceptance rate, obtained with the fourth-order leap-frog algorithm for the low resolution (128$^3$) runs. The global step-size is given by $\Delta\tau_{\rm 4th}=(2i-(2i)^{1/3})m\epsilon$ for $m=1,2,4,6,8,10$, with $\epsilon=0.06$.}}
\label{table_128}
\end{table}

To determine the iteration at which the HMC  sampler reaches convergence, we compare the power spectrum of a specific iteration with {\color{black} a reference converged power spectrum. The latter is  obtained with the second-order leap-frog algorithm, from computing the average power spectrum from iteration $3000$ to $12000$, i.e., taking samples well after the chain has passed the burn-in phase.} 
We estimate that convergence has set in when the ratios  between the reference power spectrum and that of a certain iteration are compatible with each other within 2.5\%.
We further assess the convergence of the chains in a robust way using the Gelman-Rubin estimator in \S \ref{sec:gelmanrubin}.

\subsubsection{Parameter study: optimal step-size and convergence}
\label{sec:param}

\begin{figure}
\centering
\includegraphics[width=8.2cm]{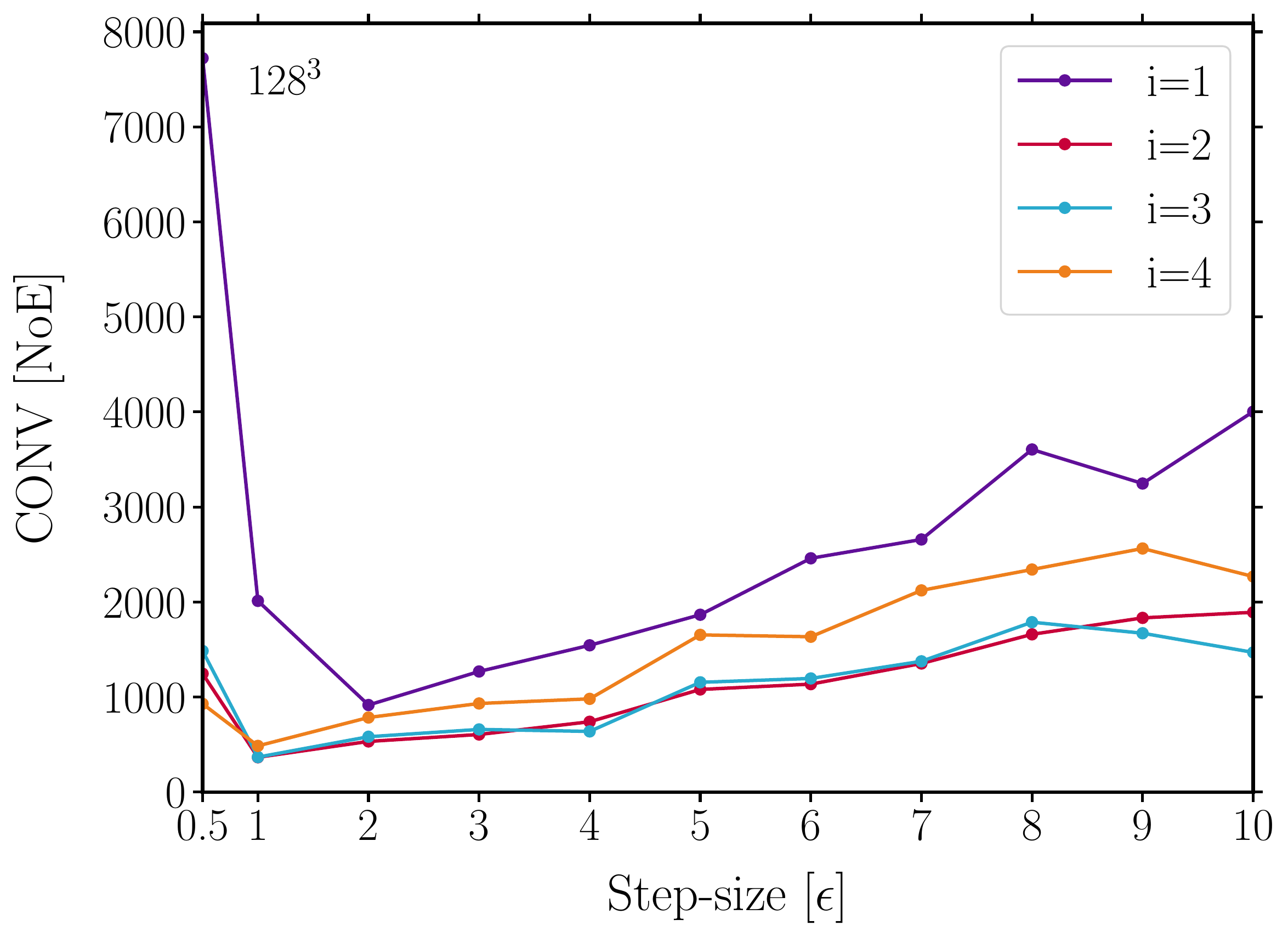}
\caption{\label{fig:time}{Number of evaluations of the Hamiltonian equations of motion required to achieve the convergence as a function of the step-size for the different values of $i$.}} 
\end{figure}

\begin{figure*}
\centering
\includegraphics[width=15.5cm]{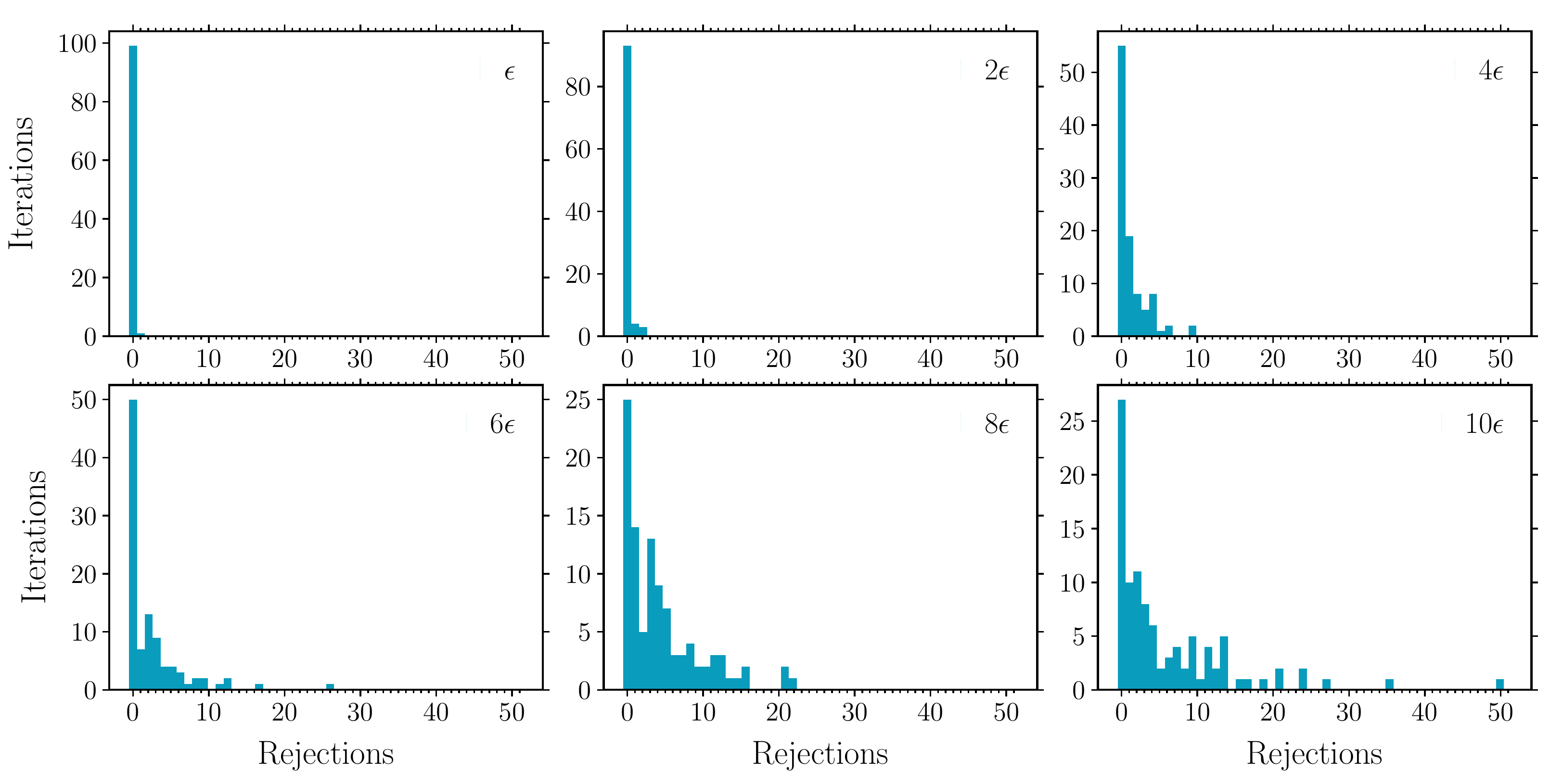}
\caption{\label{fig:hist} Number of rejected iterations for $i=1$ and the different values of step-size for the 128$^3$ runs in the fourth-order scheme.} 
\end{figure*}

\begin{figure*}
\centering
\includegraphics[width=16cm]{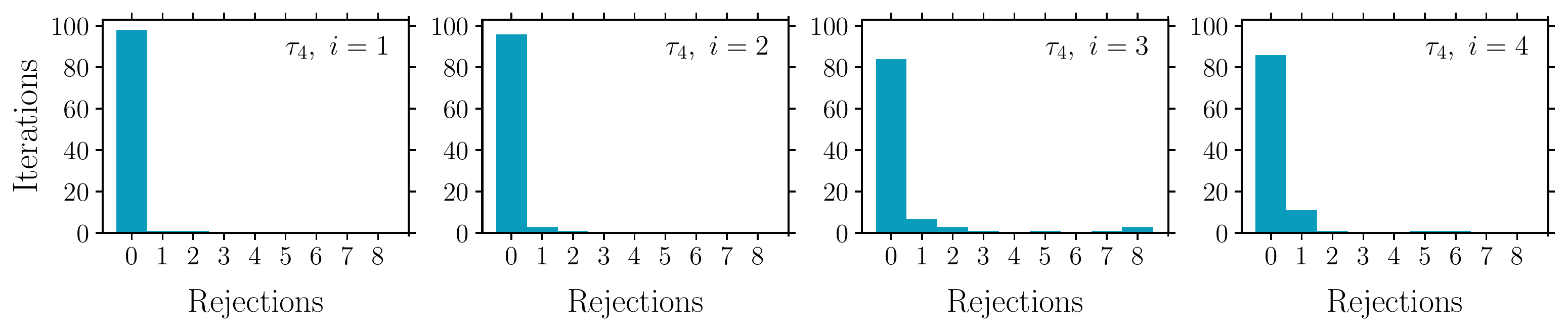}
\caption{\label{fig:histi} {Number of rejected iterations for a step-size of $\epsilon$ and the different values of $i$ in the fourth-order scheme. From the left to the right: $i=1,2,3$ and $4$  for the 128$^3$ runs.}}
\end{figure*}

\begin{figure*}
\centering
\includegraphics[width=8.1cm]{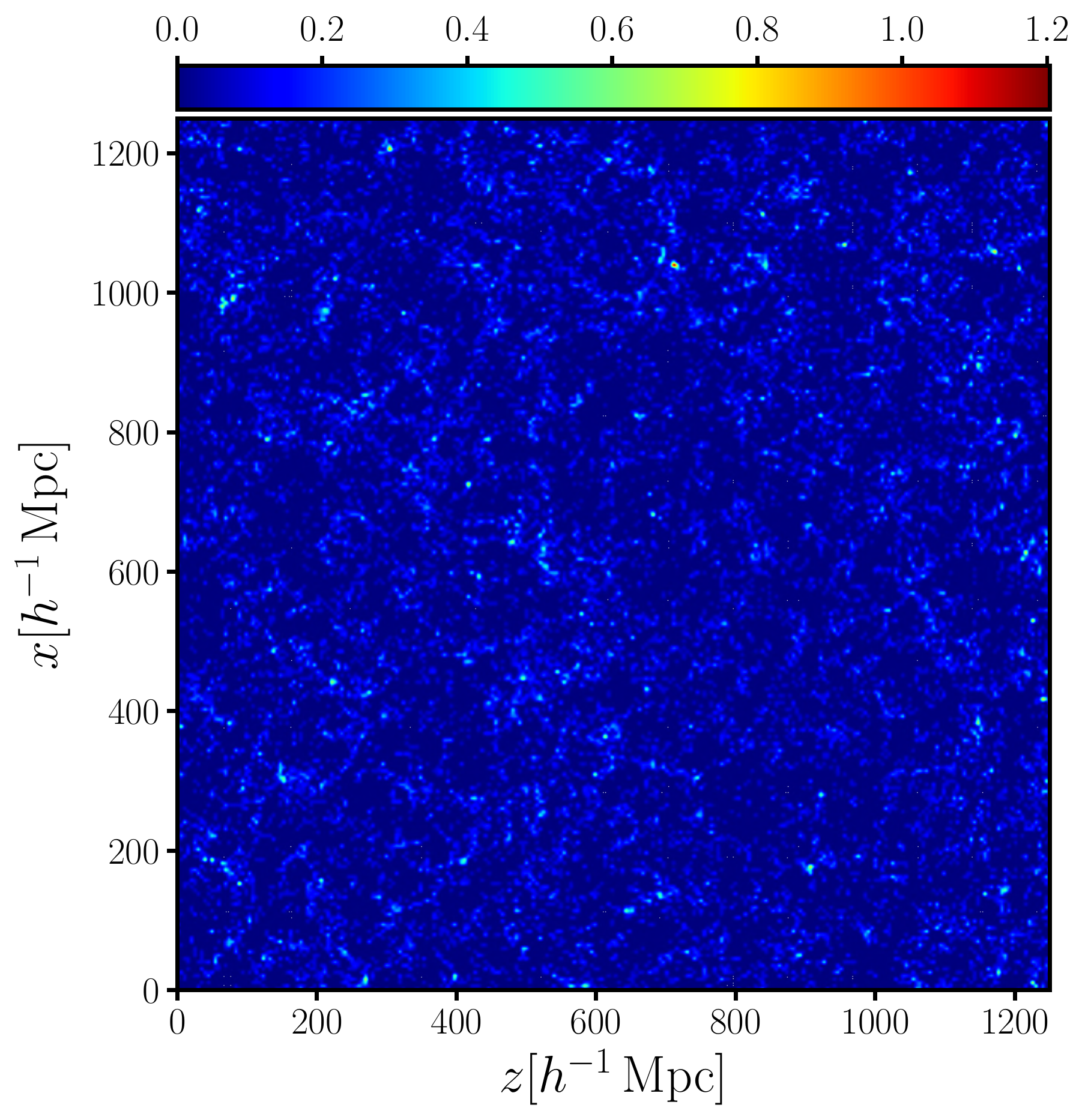}
\includegraphics[width=8.1cm]{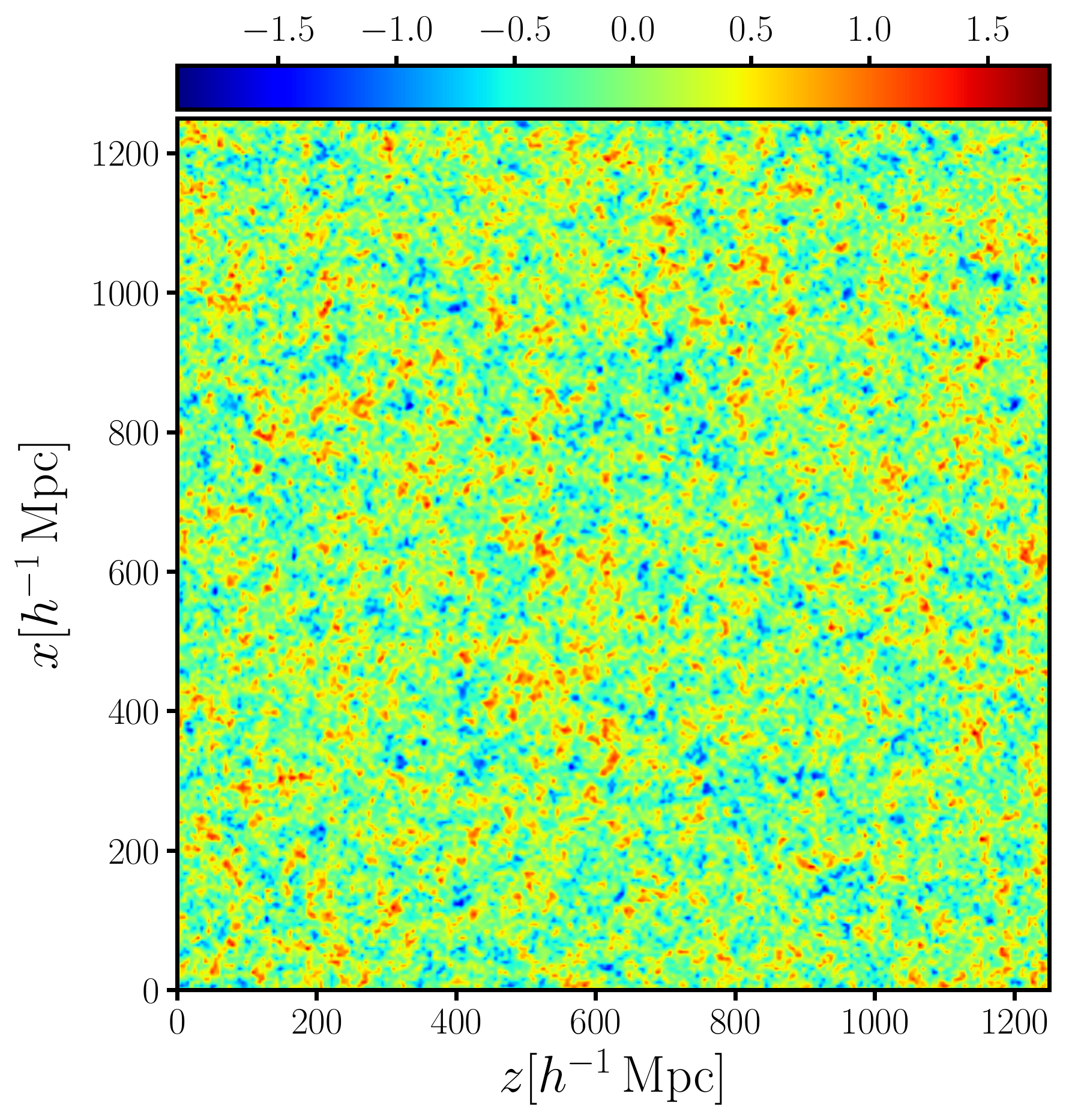}
\caption{\label{fig:final_birth} Comparison between the halo density filed (left) and the reconstructed primordial fluctuations, $\delta(x)$, with fourth-order leap-frog algorithm, for $i=3$ and a step-size $\epsilon=0.06$ (right). We have taken a volume of $(1250$ $h^{-1}$ Mpc$)^{3}$ and $256^{3}$ cells. The slice was obtained integrating $10$ cells in $y$ direction,  corresponding to a thickness of $\sim50$  $h^{-1}$ Mpc. 
}
\end{figure*}

To define our reference computation we start with the second-order leap-frog algorithm, and determine an optimal  step-size of $\epsilon=0.06$, multiplied by a uniform random number with an additionally drawn random number of steps in each iteration, $N_{\rm eval}$, shown in Eq. \ref{eq:neval} \citep[see][and \S \ref{sec:second}]{Neal}. 

For the fourth-order discretisation scheme, the optimal setup will be investigated in the following subsections.
We start with the low resolution case, which permits us to scan more broadly the parameter space.

\begin{itemize}
 
 \item Low resolution studies: 128$^3$ cells

This study has been done for different number of steps $i$ within one iteration, and for different step-sizes (multiple values of $\epsilon$), to analyze the convergence, the computation time, and the acceptance rate. This last parameter expresses the percentage of iterations that have been accepted at the first time. 
A too large step-size will result in a very low acceptance rate for the new states, and a too small step-size can waste computation time or will lead to a slow exploration of the parameter space.  

Table \ref{table_128} shows the iteration at which the chain converges, the number of evaluations of the Hamilton's equations of motion required to achieve that convergence, and the acceptance rate. This has been computed for each value of $i$ and different step-sizes. To suppress  the dependence on the starting point of the chain, all runs have been  performed for $5$ different seeds,  yielding stable results, as can be seen in the small fluctuations in figure \ref{fig:time}. Hence, the results in table \ref{table_128} represent the average over the $5$ chains. 
 {The highest acceptance rates are obtained for a step-size of $0.5 \epsilon$ in all four cases of $i$. However, the convergence is achieved at a higher iteration than for the other configurations. For a step-size of $\epsilon$ the acceptance rate is still high, and the iteration of convergence has  significantly decreased with respect to the previous case. Thus, for the cases of $i = 2$, $i=3$ and $i=4$ this is the optimal configuration. For the case of $i=1$, table \ref{table_128} shows that the optimal step-size value is the one of $2\epsilon$, for which the convergence is reached at iteration $250$ with a relative high acceptance rate, while for a step-size of $\epsilon$ we need $650$ iterations to converge.}  However, as we increase the step-size value, we can observe that convergence is achieved  at a similar number of iterations as in the case of  $2\epsilon$, but with at the expense of a higher computational cost due to the number of rejected samples.
We also find that, for larger  values of $i$, the acceptance ratio decreases faster with increasing step-sizes, with some exceptions for $i=4$, which implies that the computation time increases. 
This can also be seen in figure \ref{fig:time}, where we have represented the convergence time over the step-size value. In particular, we find a linear positive slope from a step-size of $\epsilon$ to $10 \epsilon$, with the exception of step-size $\epsilon$ for case $i=1$. In this case, the global transformation to a new state, including the backward step, is presumably too short to take advantage of the fourth-order discretisation (see discussion at the end of \S \ref{sec:higher}). {This effect can also be observed taking a step-size of $0.5 \epsilon$, especially for $i=1$, which implies a significant increase of the number of evaluations of the Hamiltonian equations of motion and, therefore, a higher computational cost.}

\begin{table*}
{ \color{black}
\begin{tabular}{|c|c|c|c|c|c|c|c|}
\hline \hline
Scheme &  \begin{tabular}[c]{@{}c@{}} $\langle$NoE$\rangle$ \end{tabular} &  \begin{tabular}[c]{@{}c@{}} CONV \\ $[$NoI$]$\end{tabular} &  \begin{tabular}[c]{@{}c@{}} CONV$^{\rm  eff}$\\ $[$NoE$]$  \end{tabular}&  \begin{tabular}[c]{@{}c@{}} IF$_{\rm CONV}$ \end{tabular}& \begin{tabular}[c]{@{}c@{}} AR$_{\rm CONV}$ \\ $[$\%$]$  \end{tabular} & \begin{tabular}[c]{@{}c@{}} $\Delta \tau_{\rm CONV}$  \\ $[\epsilon]$\end{tabular} &  \begin{tabular}[c]{@{}c@{}} $\Delta \tau^{\rm eff}_{\rm CONV}$\end{tabular} 
\\ \hline
\hline
\multicolumn{8}{|c|}{2nd order}\\\hline
\rowcolor{gray!30}{\begin{tabular}[c]{@{}c@{}}$\mathcal{T}_2$\\ $N_{\rm eval}=2$ \end{tabular}} 
& 1.5   & 4320  & 12557  & 28.22 & 61.11& 0.68 & 0.277  \\  \hline
\rowcolor{gray!30}{\begin{tabular}[c]{@{}c@{}}$\mathcal{T}_2$\\ $N_{\rm eval}=5$ \end{tabular}} 
& 3   & 2175  & 14833  & 33.33 & 46.96 & 0.83 & 0.130\\  
\hline
{\begin{tabular}[c]{@{}c@{}}$\mathcal{T}_2$\\$N_{\rm eval}=15$  \end{tabular}} 
& 8  &  665  & 17372 & 39.04 &  33.83 & 1.34 & 0.057\\    \hline
{\begin{tabular}[c]{@{}c@{}}$\mathcal{T}_2$ \\ $N_{\rm eval}=30$ \end{tabular}} & 15.5 & 405  & 27459  &  61.71 & 23.78 & 3.87 & 0.059\\ 
\hline
{\begin{tabular}[c]{@{}c@{}}$\mathcal{T}_2$ \\ $N_{\rm eval}=50$ \end{tabular}} & 25.5 &    242   & 33069 & 74.31 &  20.82 & 4.85 & 0.039\\ 
\hline
{\begin{tabular}[c]{@{}c@{}}$\mathcal{T}_2$ \\ $N_{\rm eval}=80$ \end{tabular}} & 40.5 &    149   & 45738 & 102.78 &  15.63 & 7.25  & 0.028\\ 
\hline
{\begin{tabular}[c]{@{}c@{}}$\mathcal{T}_2$ \\ $N_{\rm eval}=100$ \end{tabular}} & 50.5 &    113   & 53789 & 120.87 &  15.46 & 7.54  & 0.023\\ 
\hline
\multicolumn{8}{|c|}{4th order}
\\
\hline
{\begin{tabular}[c]{@{}c@{}}$\mathcal{T}_4$\\ $i=1$ \end{tabular}} & 3 &    340& 1496 & 3.36 &  69.85 & 1.48  & 0.35\\ 
\hline
{\begin{tabular}[c]{@{}c@{}}$\mathcal{T}_4$\\ $i=2$ \end{tabular}} & 5&    100& 608 & 1.37 &  83.79  & 2.41 &  0.40\\ 
\hline
\rowcolor{gray!30}{\begin{tabular}[c]{@{}c@{}}$\mathcal{T}_4$\\ $i=3$ \end{tabular}} & 7&    33   & 445 & ref &  71.30 & 4.18  & 0.43\\   
\hline
{\begin{tabular}[c]{@{}c@{}}$\mathcal{T}_4$\\ $i=4$ \end{tabular}} & 9&    33   & 504 & 1.13 &  67.13 & 6.00  & 0.45 
\\\hline
\multicolumn{8}{|c|}{4th order with random $i$}
\\ \hline
\rowcolor{gray!30}{\begin{tabular}[c]{@{}c@{}}$\mathcal{T}_4$\\ $i\in\{1,2,3,4\}$ \end{tabular}} & 6 &  41 & 349 & 0.78 & 75.54 & 3.27  & 0.42\\  
\hline
{\begin{tabular}[c]{@{}c@{}}$\mathcal{T}_4$\\ $i\in\{2,3,4,5,6\}$ \end{tabular}} & 9 &  25 & 439 & 0.99 & 60.03 & 5.22  & 0.35\\  
\hline
\end{tabular}
}
\caption{{ \color{black} Comparison between the second  and fourth-order leap-frog schemes ($\mathcal{T}_2$ and $\mathcal{T}_4$, respectively), for the high resolution ($256^{3}$) runs.
The 1st column indicates the specific settings of the scheme, the 2nd one the average number of evaluations of Hamilton's equations of motion (NoE), the 3rd column shows the iteration of convergence, and the 4th one the number of evaluations until convergence (taking into account the rejections). In the 5th column we have the improvement factor (IF) of each scheme vs the fastest $\mathcal{T}_2$ in convergence (CONV) and the 6th one the acceptance rate (AR$_{\rm CONV}$) until  convergence.
The  $\mathcal{T}_4$  with $i=1$ was run with a basic step-size of  $2\epsilon$, while the rest used $\epsilon$ instead, following the analysis shown in figure \ref{fig:time}. This results in global step-sizes given by equations    \ref{eq:Tau2ndorder} and \ref{eq:Tau4thorder} indicated in the 7th column. The last one shows the effective global step-size weighted with the respective AR$_{\rm CONV}$ (equation \ref{eq:effective_tau}). The $N_{\rm eval}=2$ case  is the one with the least number of evaluations, which converges. One might be cautious about this case, given the risk of resonant trajectories for such a low number of evaluations ($\langle$ NoE$\rangle=1.5$). In fact, the $N_{\rm eval}=1$ case does not converge in general. The best second and fourth-order cases are highlighted in gray. We choose the best fourth-order case with a fixed number of evaluations as the reference (ref).}}
\label{table_conv}
\end{table*}

\begin{figure*}
\centering
\includegraphics[width=13.5cm]{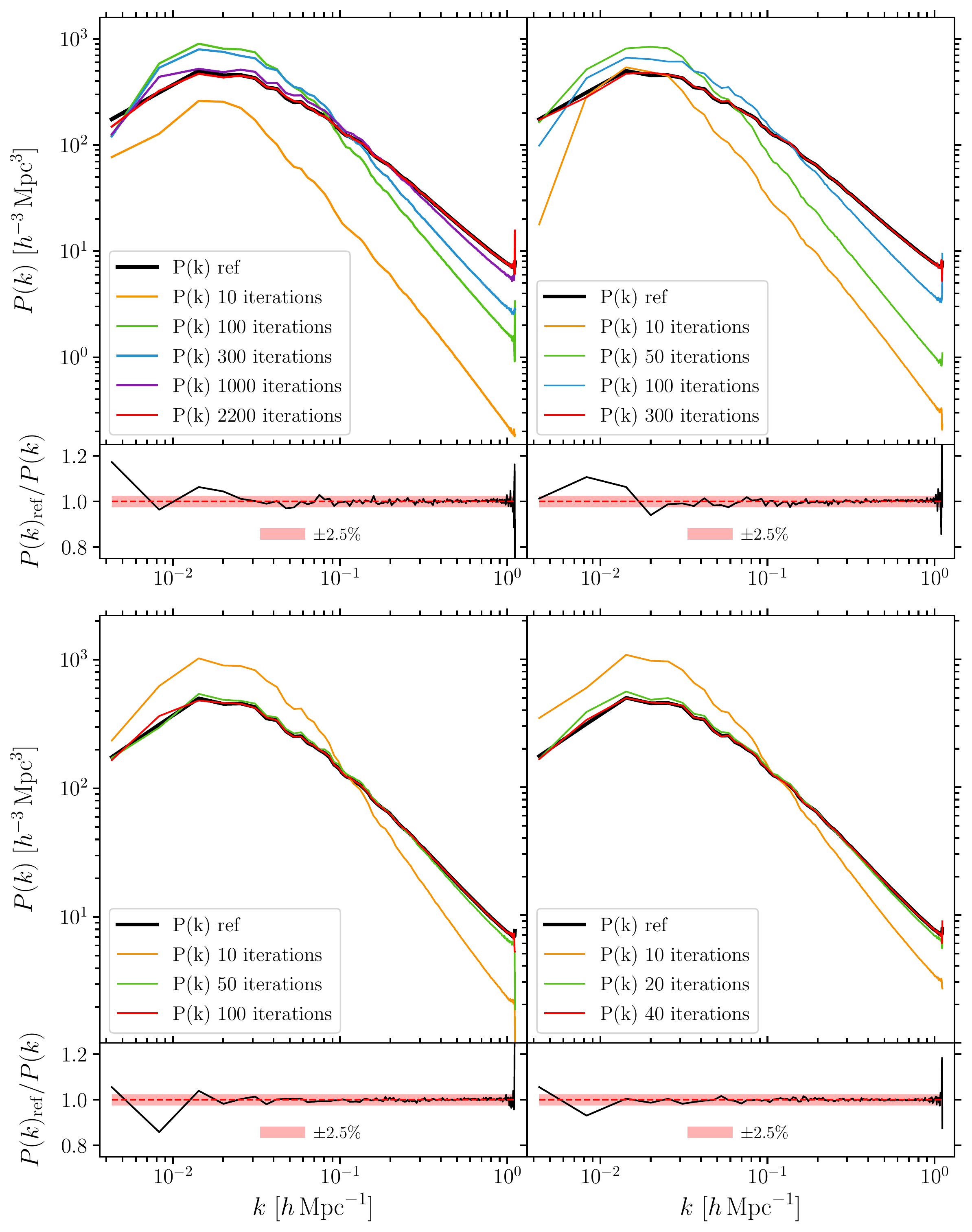}
\put(-237,465){$\mathcal{T}_2$, $N_{\rm eval}=5$}
\put(-48,465){$\mathcal{T}_4$, $i=1$} 
\put(-48,229){$\mathcal{T}_4$, $i=3$}
\put(-218,229){$\mathcal{T}_4$, $i=2$} 

\caption{\label{fig:compare} Power spectra for different iterations (coloured lines) compared to the reference {\color{black} averaged  (over 9000 samples)} converged power spectrum (black line) for the high resolution (256$^3$) runs. The lower panels show the ratio between the converged sample for each setup with respect to the reference converged sample. The power spectrum represented with the red line corresponds to that at iteration of convergence. The subplots show the ratio between the reference converged power spectrum and the converged one for each setting. {\bf Upper left}: power spectrum for different iterations with the second-order leap-frog algorithm. {\bf Upper right}: power spectrum for different iterations with the fourth-order leap-frog algorithm, for $i=1$ and a step-size $2\epsilon$. {\bf Lower left}: power spectrum for different iterations with the fourth-order leap-frog algorithm, for $i=2$ and a step-size $\epsilon$. {\bf Lower right}: power spectrum for different iterations with the fourth-order leap-frog algorithm, for $i=3$ and a step-size $\epsilon$. }
\end{figure*}

\begin{figure*}
\centering
\includegraphics[width=15.5cm]{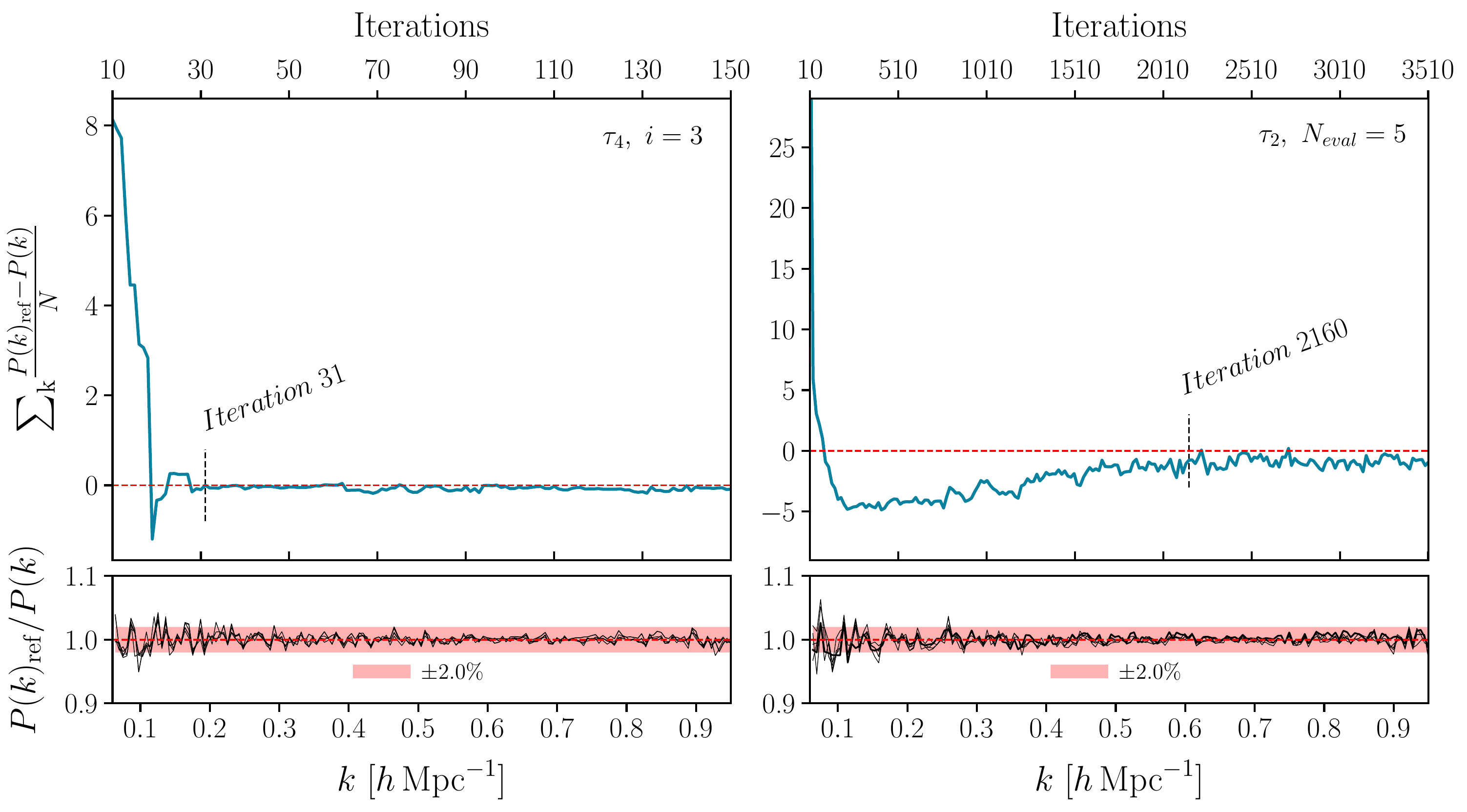}
\caption{\label{fig:diffpk}{\color{black} Difference of the power spectrum of each iteration and the reference one, summing up for all modes, $0.06<k<0.95$ $h$ Mpc$^{-1}$ to avoid cosmic variance. {\bf The panel on the left} shows the fourth-order scheme, $\mathcal{T}_4, i=3$ and {\bf the panel on the right} the second-order one, $\mathcal{T}_2, N_{\rm eval}=5$. {\bf The lower panels} show the ratio of the converged power spectrum and the consecutive ones, with the black lines}}
\end{figure*}

Figure \ref{fig:hist} represents the acceptance for the case $i=1$ as a function of the step-size $m\epsilon$. We can see that, as the step-size increases, the number of rejections becomes larger. For the case of a step-size of $\epsilon$, $97.0\%$ of the iterations are accepted at the first trial, and $3.0\%$ at the second one,  i.e. with only one rejection. For the case of $2\epsilon$, we can observe that there is increment, although  small number of iterations that are accepted at the second and third time. The histogram in the lower right panel of figure \ref{fig:hist}, for a step-size value of $10\epsilon$, shows that iterations can be rejected up to $50$ times before being accepted, which dramatically increases the computational cost. The same behaviour has  been found for  cases  $i=2$, $i=3$ and $i=4$ as it is shown in table \ref{table_128}.

Figure \ref{fig:histi} shows the acceptance for a step-size of $\epsilon$ for the four studied values of $i$. We can see in the panel on the left that, for the case of $i=1$, almost all iterations are accepted without rejections. For $i=2$ we find that there is a very high acceptance ratio at the first trial, but some rejections start to appear after one trial. For $i=3$ the number of rejections increase, although remaining low. {Finally, for $i=4$ we find, in the panel of the right, a similar behaviour to  the previous case, decreasing the number of rejections at high trials.}
\\
 \item High resolution studies: 256$^3$ cells.

Once we have studied the low resolution case, we can now focus on fewer configurations at a higher resolution. 
{For the second-order leap-frog algorithm we investigate, first, the optimal $N_{\rm eval}$ which goes into Eq. \ref{eq:neval}. Previous studies at lower resolution ($128^3$ cells) showed an optimal step-size value of $\epsilon$ for different configurations, i.e,  different $N_{\rm eval}$ values. Hence,  we have chosen this step-size in all the cases of this scheme shown in table \ref{table_conv}. In the fourth-order method we present the results for the configurations $2\epsilon$, for $i=1$ and $\epsilon$ for $i=2$, $i=3$ and $i=4$. These were the most efficient step-sizes for each value of $i$, as we could see in table \ref{table_128} and figure \ref{fig:time}. 
In table \ref{table_conv}, we can compare the second and  fourth-order leap-frog schemes at a higher resolution of $256^3$ cells. Each value of the table is an average over $4$ different seeds. We have empirically  found that for our setting this is a reasonable number to avoid being much affected by the initial conditions. In fact, in a number of relevant cases, several measures have not changed, such as the number of evaluations or the correlation length for the fourth-order $i=3$ case (see \S \ref{sec:corr}).
The table shows seven $N_{\rm eval}$ values from $2$ to $100$ in the second-order case, and six configurations for the fourth-order method: four different $i$ values from $1$ to $4$, and the last two cases of the table, with a random value of $i$ in each iteration, from $1$ to $4$ and from $2$ to $6$, respectively.
Table \ref{table_conv} shows the average number of evaluations of Hamilton's equations of motion required in each iteration ($\langle \rm{NoE} \rangle$), the iteration of convergence (NoI) and the number of evaluations until that convergence, taking into account the rejections (NoE). From this study we obtain the ratio of the number of evaluations until  convergence is reached between each case and that of the reference (IF). We can also find the acceptance rate until the convergence (AR) and the global step-sizes given by Eqs.
~\ref{eq:hrel} and \ref{eq:Tau4thorder}, with the effective global step-size weighted with respect to the acceptance rate as follows:}
  \color{black}{
  \ba
  \Delta \tau^{\rm eff}\equiv\frac{\Delta \tau\cdot {\rm AR}}{\langle{\rm NoE}\rangle}\,.
  \label{eq:effective_tau}  \ea
  This quantity is defined based on the time step-size $\Delta\tau$, penalised by the acceptance rate AR, and the number of evaluations $\langle{\rm NoE}\rangle$. 
}

{We have chosen as the fourth-order reference case, the one of $i=3$, which is the most efficient one in terms of convergence from the cases $i=1,2,3$ and $4$. 
We note that the fourth-order case has a random step-size, following equation   \ref{eq:Tau4thorder}, but the number of evaluations is in general fixed. For this reason, we tested also the behaviour of the fourth-order scheme with random number of evaluations (randomizing $i$) as shown in the last two rows of table \ref{table_conv}. 
 The convergence and correlation length (shown in \S \ref{sec:corr}) are not significantly better to that one of $i=3$. Thus, moderate improvements could be achieved, by considering more cases, than the ones computed in this study.
 For the second-order method, represented in the seven first rows of the table, we find an optimal configuration in terms of convergence for the lowest number of evaluations  of $N_{\rm eval}=2-5$. This is however, a factor $\sim30$ less efficient than the best higher fourth-order case. 
The $N_{\rm eval}=2$ case with $\langle$NoE$\rangle=1.5$ has converged considering different seeds, however, for  $N_{\rm eval}=1$, we did not find stable convergence, as there is a high risk of producing resonant trajectories.}

A visual impression of the reconstruction is shown in figure \ref{fig:final_birth}, where the input catalog and the corresponding reconstruction of the linear density field using the fourth-order discretisation scheme are shown. Here we  can qualitatively verify that the discrete number counts of objects on the left panel is translated into a continuous density field on the right panel. This is essential to primordial density reconstructions, as we need to obtain a clean Gaussian field on which we can make non-linear cosmic evolution operations \citep[see e.g. scheme in][relating the Gassian field to the final galaxy distribution, and the corresponding power spectra]{Kitaura2013}.

Figure \ref{fig:compare} shows the convergence of the power spectra as a function of the number of iterations for the optimal case {of the second-order  leap-frog algorithm ($\mathcal{T}_2, N_{\rm eval}=5$) and for the cases of $i=1,2$ and $3$ of the fourth-order scheme}.  While in the second-order method convergence is reached at iteration $\sim 2200$, with the fourth-order scheme we  get converged samples at iteration $\sim 280$ for the case $i=1$, $\sim 100$ for $i=2$, and $\sim 40$ for $i=3$, as we can read from table \ref{table_conv}. Note that figure \ref{fig:compare} shows results for a particular seed, so that the optimal values can vary with respect to the ones presented in the table, as the average over different seeds. {The ratio between the converged reference power spectrum, represented with the black line, and the power spectrum at the estimated convergence iteration, are shown in the lower panels of figure \ref{fig:compare}. We find that the ratios are compatible within a $2.5\%$ error, represented with the red band.} 

{To further assess  the convergence of the second and fourth-order leap-frog algorithms, we have performed an additional analysis. In particular, we compute,  for each iteration, the difference of the individual power spectrum and the reference one, summing up for all modes, $k$, as represented in figure \ref{fig:diffpk}. Convergence is achieved when this difference is smaller than a threshold (in this case, the criterion is $<1$) for at least $10$ consecutive iterations. As this is for one seed, we can see the convergence iterations are in agreement with figure \ref{fig:compare} and table \ref{table_conv}. {\color{black} It is also remarkable to observe that the modes averaged difference between the reference  and sampled power spectra at different iterations is always considerably closer to zero for the fourth-order case than for the second-order one}. Then, we represent the ratio of the reference power spectrum with the power spectrum at that iteration (previously estimated), and the next ones, where we find that they are compatible with each other within $2\%$. This test is reassuring, as both convergence criteria are in excellent agreement.}
 
Having set the number of forward steps to $i=3$, and the step-size to $\epsilon$, which is the most optimal configuration, we can now proceed to study the convergence of the fourth-order leap-frog algorithm compared to the second-order one in {\color{black} an additional} robust way. 

\end{itemize}

\subsubsection{Robust convergence assessment: Gelman-Rubin test}
\label{sec:gelmanrubin}

\begin{figure}
\centering
{\includegraphics[width=7.8cm]{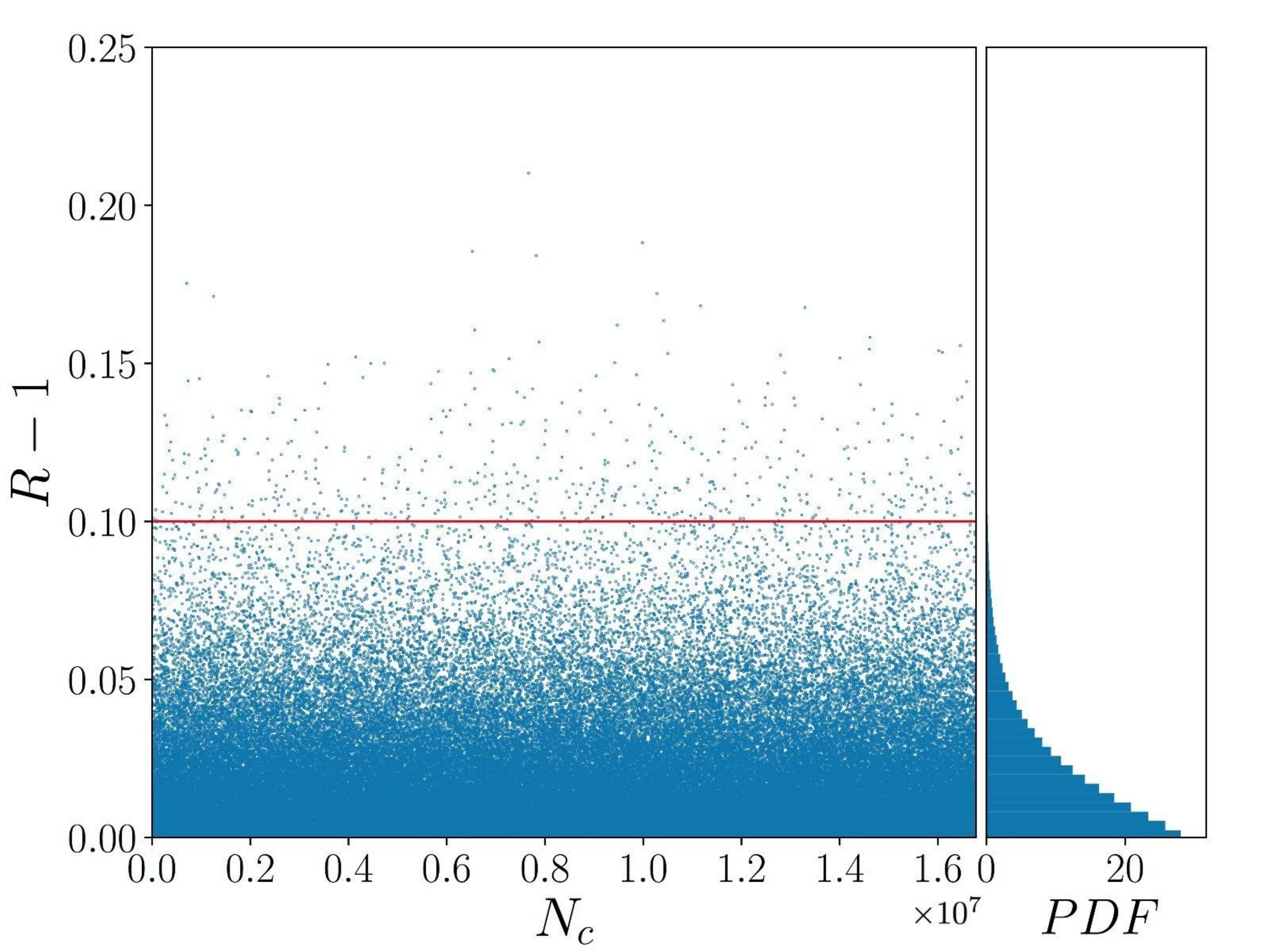}}
\put(-21,35){$\mathcal{T}_4$}\\
{\includegraphics[width=7.8cm]{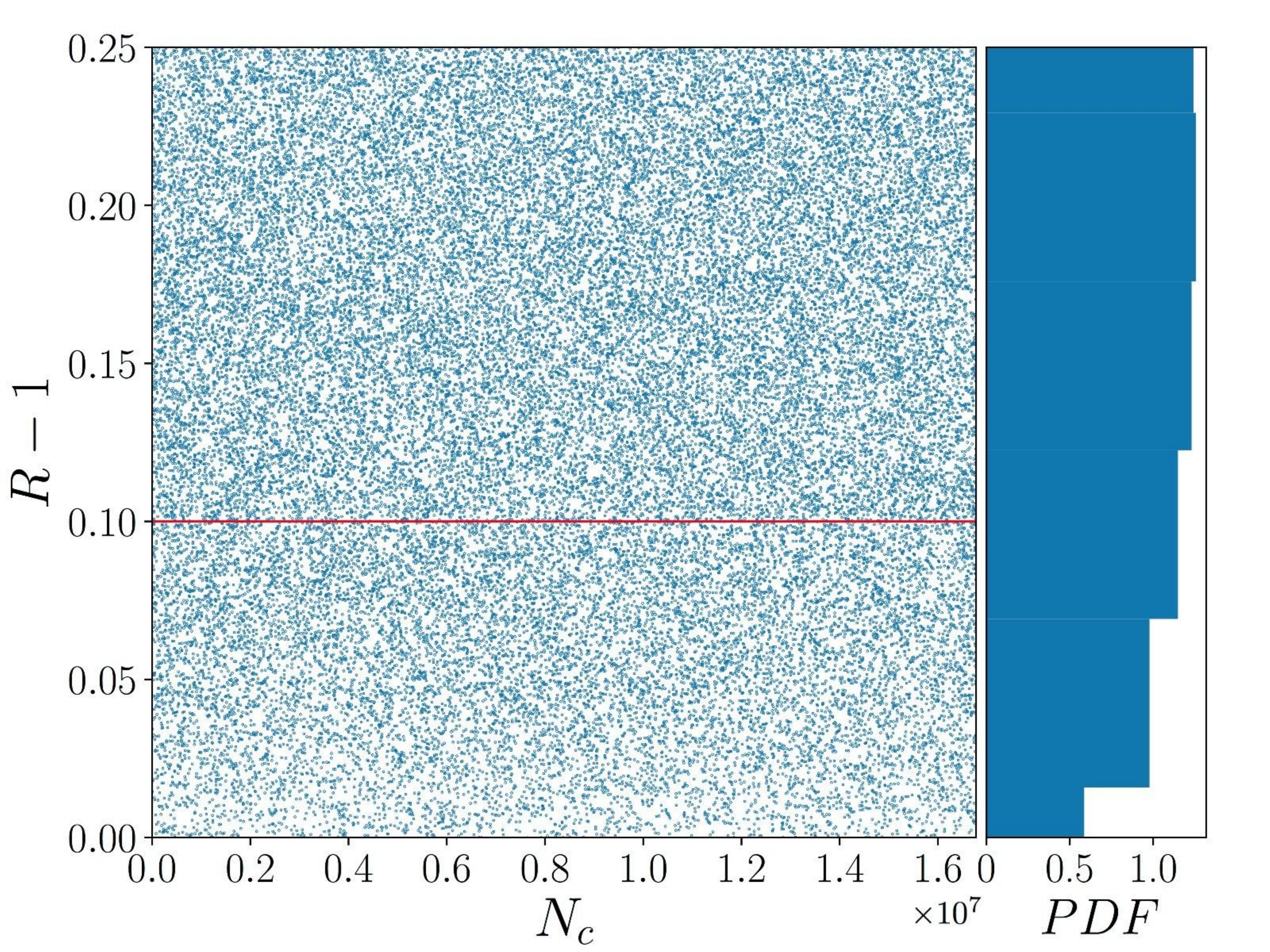}}
\put(-21,35){$\mathcal{T}_2$}\\
{\includegraphics[width=7.8cm]{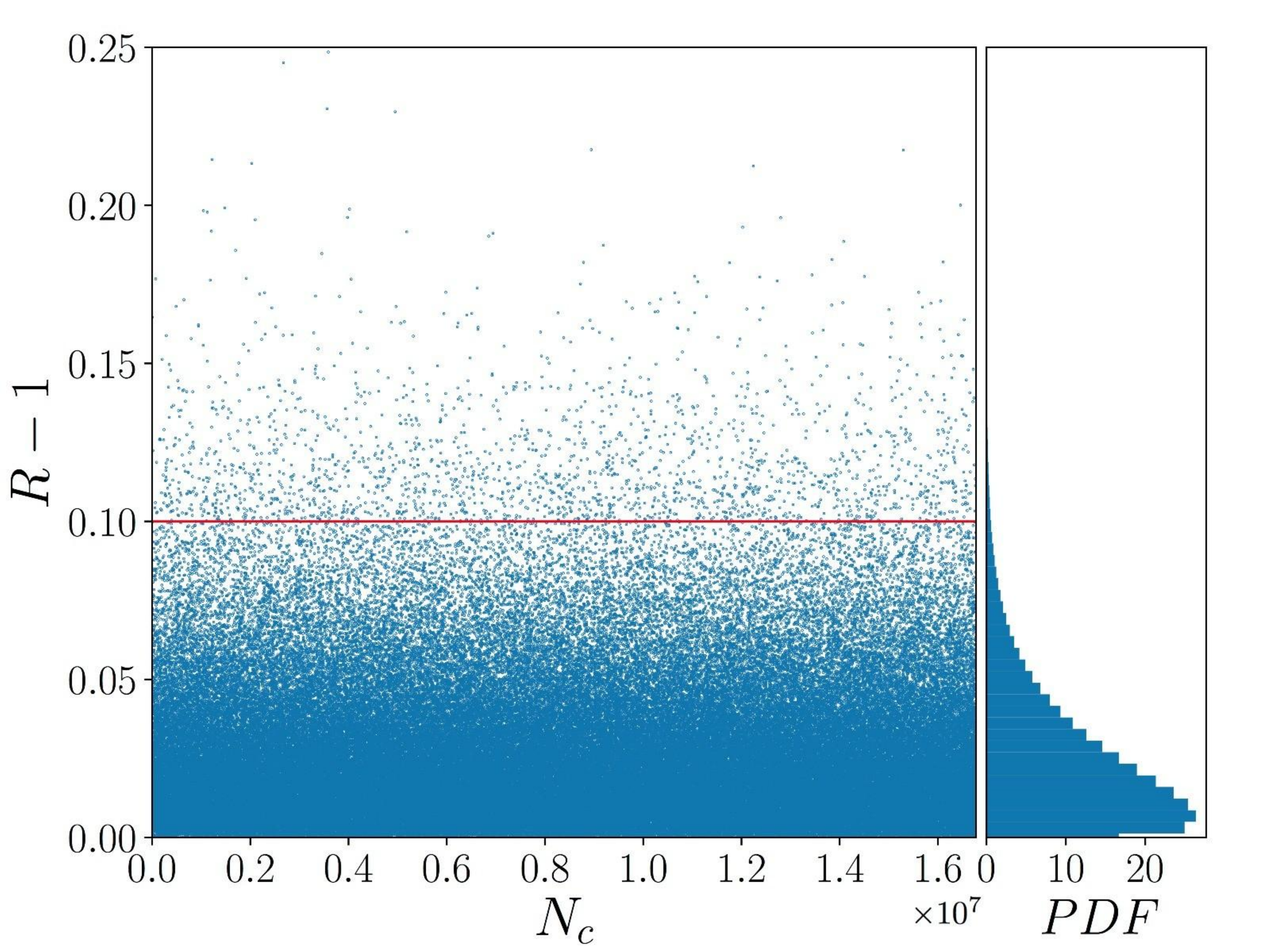}}
\put(-21,35){$\mathcal{T}_2$}
\caption{\label{fig:gr} {\color{black}{\bf Upper panel:} Gelman-Rubin test from $40$ to $500$ iterations with the best fourth-order leap-frog algorithm, for the configuration $i=3$. {\bf Middle panel:} Gelman-Rubin test from $3000$ to $3460$ iterations with second-order leap-frog algorithm  with $\langle$NoE$\rangle=5.5$. {\bf Lower panel:} Gelman Rubin test from $3000$ to $12000$ iterations for second-order leap-frog algorithm with $\langle$NoE$\rangle=5.5$. The red line represents the $R-1=0.1$ parameter.}} 
\end{figure}

\begin{table*}
{ \color{black}
\begin{tabular}{|c|c|c|c|c|c|c|c|c|}
\hline \hline
scheme &  \begin{tabular}[c]{@{}c@{}} $\langle$NoE$\rangle$ \end{tabular} & \begin{tabular}[c]{@{}c@{}} CL \\ $[$NoI$]$\end{tabular} &  \begin{tabular}[c]{@{}c@{}} CL$^{\rm eff}$
\end{tabular}&   \begin{tabular}[c]{@{}c@{}} IF$_{\rm CL}$ \end{tabular}&  \begin{tabular}[c]{@{}c@{}} AR$_{\rm CL}$ \\ $[$\%$]$  \end{tabular} & \begin{tabular}[c]{@{}c@{}} NoIS
   \end{tabular}&  \begin{tabular}[c]{@{}c@{}} $\Delta \tau_{\rm CL}$\\  $[\epsilon]$\end{tabular} &  \begin{tabular}[c]{@{}c@{}} $\Delta \tau^{\rm eff}_{\rm CL}$\end{tabular} 
\\ \hline
\hline
\multicolumn{9}{|c|}{2nd order}\\ \hline
{\begin{tabular}[c]{@{}c@{}}$\mathcal{T}_2$\\ $N_{\rm eval}=2$ \end{tabular}} 
& 1.5   & 290  & 527  & 3.66 & 80.29  & 75 & 0.76 & 0.407  \\  \hline
{\begin{tabular}[c]{@{}c@{}}$\mathcal{T}_2$\\ $N_{\rm eval}=5$ \end{tabular}} 
& 3   &  200 & 826 & 5.13  & 74.33 & 89 & 1.49 & 0.369 \\  \hline

{\begin{tabular}[c]{@{}c@{}}$\mathcal{T}_2$\\$N_{\rm eval}=15$  \end{tabular}} 
& 8   & 70 & 904 & 5.61 &  60.37& 105 & 4.00 & 0.302   \\    \hline
{\begin{tabular}[c]{@{}c@{}}$\mathcal{T}_2$ \\ $N_{\rm eval}=30$ \end{tabular}} & 15.5  & 22 & 609 & 3.78 & 60.39 & 167 & 7.69 &   0.300   \\ \hline
{\begin{tabular}[c]{@{}c@{}}$\mathcal{T}_2$ \\ $N_{\rm eval}=50$ \end{tabular}} & 25.5  & 9  & 433 &  2.69 & 59.09& 203 & 12.42 & 0.288
\\ \hline
\rowcolor{gray!30}{\begin{tabular}[c]{@{}c@{}}$\mathcal{T}_2$ \\ $N_{\rm eval}=80$ \end{tabular}} & 40.5  & 4  & 369 & 2.29 & 55.38& 281 & 19.61  & 0.268 \\ \hline
{\begin{tabular}[c]{@{}c@{}}$\mathcal{T}_2$ \\ $N_{\rm eval}=100$ \end{tabular}} & 50.5  & 4  & 390 & 2.42 &  57.40 & 331 & 24.23  & 0.227\\ 
\hline
\multicolumn{9}{|c|}{4th order}
\\ \hline
{\begin{tabular}[c]{@{}c@{}}$\mathcal{T}_4$\\ $i=1$ \end{tabular}} & 3   & 123 & 2000 & 12.42 & 41.81 & 7 & 1.48  &  0.206   \\ \hline
{\begin{tabular}[c]{@{}c@{}}$\mathcal{T}_4$\\ $i=2$ \end{tabular}} & 5 & 55 & 474 & 2.94 &  83.20 & 1  & 2.41 &   0.401  \\ \hline
\rowcolor{gray!30}{\begin{tabular}[c]{@{}c@{}}$\mathcal{T}_4$\\ $i=3$ \end{tabular}} & 7 & 18  & 161 & ref &  83.53& ref & 4.18  & 0.498 \\   \hline
{\begin{tabular}[c]{@{}c@{}}$\mathcal{T}_4$\\ $i=4$ \end{tabular}} & 9 & 16  & 163 & 1.01 &   84.5& 0.4 & 6.12  & 0.575 \\\hline
\multicolumn{9}{|c|}{4th order with random $i$}
\\ \hline
\rowcolor{gray!30}{\begin{tabular}[c]{@{}c@{}}$\mathcal{T}_4$\\ $i\in\{1,2,3,4\}$ \end{tabular}} & 6  & 28 & 214 & 1.33 & 82.48 & -0.6 & 3.32  & 0.456   \\  \hline
{\begin{tabular}[c]{@{}c@{}}$\mathcal{T}_4$\\ $i\in\{2,3,4,5,6\}$ \end{tabular}} & 9 & 14 & 263 & 1.63 & 73.54& -0.04 & 5.43  & 0.443   \\  \hline
\end{tabular}
}

\caption{{ \color{black} Comparison between the second and fourth-order leap-frog schemes ($\mathcal{T}_2$ and $\mathcal{T}_4$, respectively), for the high resolution ($256^{3}$) runs.
The 1st column indicates the specific settings of the scheme, the 2nd one the average number of evaluations of Hamilton's equations of motion (NoE), the 3rd one the correlation length in terms of iterations (NoI), and the 4th one the effective correlation length (CL$^{\rm eff}$) in units of NoE (taking into account the rejections).
The 5th column indicates the improvement factor  of each scheme in producing independent samples after convergence vs the $\mathcal{T}_4, i=3$ configuration. The 6th column shows the acceptance rate (AR) after the convergence and 7th one indicates the number of independent samples (NoIS) produced with $\mathcal{T}_4, i=3$ until each scheme converges.
The  $\mathcal{T}_4$  with $i=1$ was run with a basic step-size of  $2\epsilon$, while the rest used $\epsilon$ instead, following the analysis shown in figure \ref{fig:time}. This results in global step-sizes given by equations    \ref{eq:Tau2ndorder} and \ref{eq:Tau4thorder} indicated in the 8th column. The last one shows the effective global step-size weighted with the respective AR, given by equation \ref{eq:effective_tau}. The last  column shows the effective time steps $\Delta\tau
^{\rm eff}$. The best second and fourth-order cases are highlighted in gray.  We choose the best fourth-order case with a fixed number of evaluations as the reference (ref).}}
\label{table_corr}
\end{table*}

To verify that convergence has been reached at iteration $\sim 30$, we perform the Gelman-Rubin test. Multiple chains are supposed to converge to some stationary distribution. Hence, comparing the mean and variance within one converged chain to the samples of independent chains, gives a tool to verify convergence of Markov chains. 
In this test we have to run $\rm N_{\rm chains}$ of length $\rm N_{\rm length}$, that are supposed to have the same target distribution, but starting at different points, so each one has a different seed. The output of the chain is represented by $x_{c,s}$, with $c\in{1,2,..., \rm N_{\rm chains}}$ and $s\in {1,2,..., \rm N_{\rm length}}$. $x$ is, in this case, the over-density ${\delta_{i}}$ of each cell. The goal is to compare the variance of the $\rm N_{\rm chain}$ means of the different chains to the mean of the variance of each individual chain. The parameter $R$ introduced in \cite{gelman}, known as the  Potential Scale Reduction Factor (PSRF),  is assumed to represent a converged chain when reaching a value of $R=1.1$.

We first calculate each chain's mean value
\ba
\overline{x}_{\rm c}=\frac{1}{N_{\rm length}}\sum_{\rm s}x_{\rm c,s}\,.
\ea
Then we calculate each chain's variance
\ba
\sigma_{\rm c}^{2}=\frac{1}{N_{\rm chains}-1}\sum_{\rm s}(x_{\rm c,s}-\overline{x}_{\rm c})^{2}\,.
\ea
Then, we determine all chain's mean
\ba
\overline{x}=\frac{1}{N_{\rm chains}}\sum_{\rm c}\frac{1}{N_{\rm length}}\sum_{\rm s}x_{\rm c,s}=\frac{1}{N_{\rm chains}}\sum_{\rm c}\overline{x}_{\rm c}\,.
\ea
The weighted mean of each chain's variance is expressed as
\ba
B=\frac{N_{\rm length}}{N_{\rm chains}-1}\sum_{\rm c}(\overline{x}_{\rm c}-\overline{x})^{2}\,,
\ea
and the average variance by
\ba
W=\frac{1}{N_{\rm chains}}\sum_{\rm c}\sigma_{\rm c}^{2}\,.
\ea
Finally, the Potential Scale Reduction Factor is defined as
\ba
R=\sqrt{\frac{N_{\rm lenght}-1}{N_{\rm length}}+\frac{N_{\rm chains}+1}{N_{\rm length}N_{\rm chains}}\frac{B}{W}}\,.
\ea
\vspace{0.1in}

We have represented the range in which the Markov chain has converged and, therefore, where the HMC has reached the target distribution. As it is mentioned before, we evolve the system with Hamilton's equations of motion. However, the initial samples do not belong to the correct target distribution, but are part of the \textit{burn-in} phase. Figure \ref{fig:gr} presents the results of the Gelman-Rubin test for the fourth-order leap-frog algorithm, as compared to the second-order one. This calculation has been done for $4$ different chains. The upper panel of figure \ref{fig:gr}, shows  that a small range of $40$ to $500$ iterations, already gets the majority of the points below the solid red line, which represents $R-1 = 0.1$.  However, for  second-order leap-frog algorithm we need a larger range to find a similar behaviour in the Gelman-Rubin test: from $3000$ to $12000$ (lower panel of figure \ref{fig:gr}). If we take the same range as for the fourth-order one, we can verify that the Markov chain is far from converged (see middle panel in figure \ref{fig:gr}).

\begin{figure}
\centering
\begin{tabular}{c}
\includegraphics[width=7.2cm]{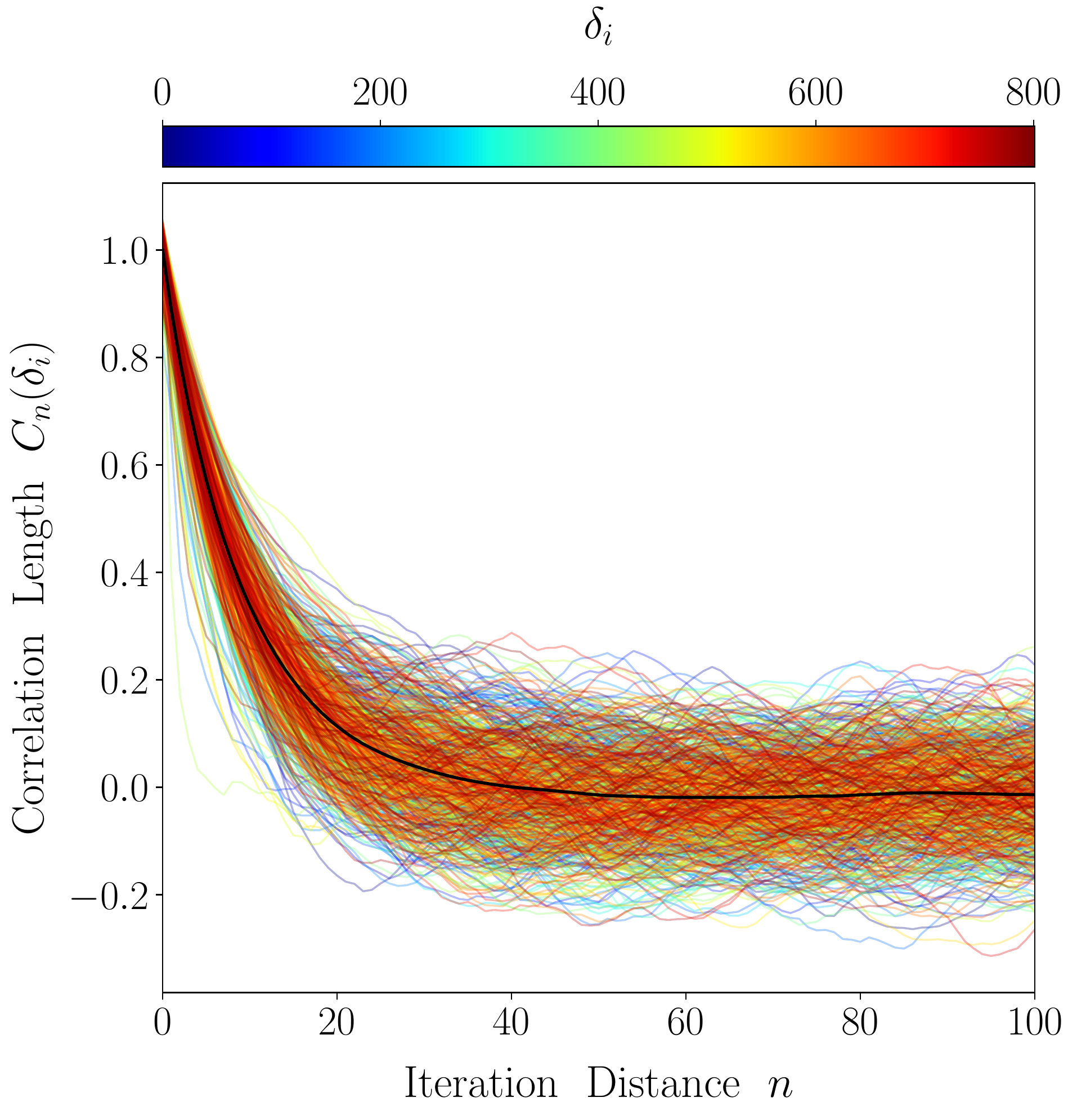}
\vspace{-0.05cm}\\
\includegraphics[width=7.2cm]{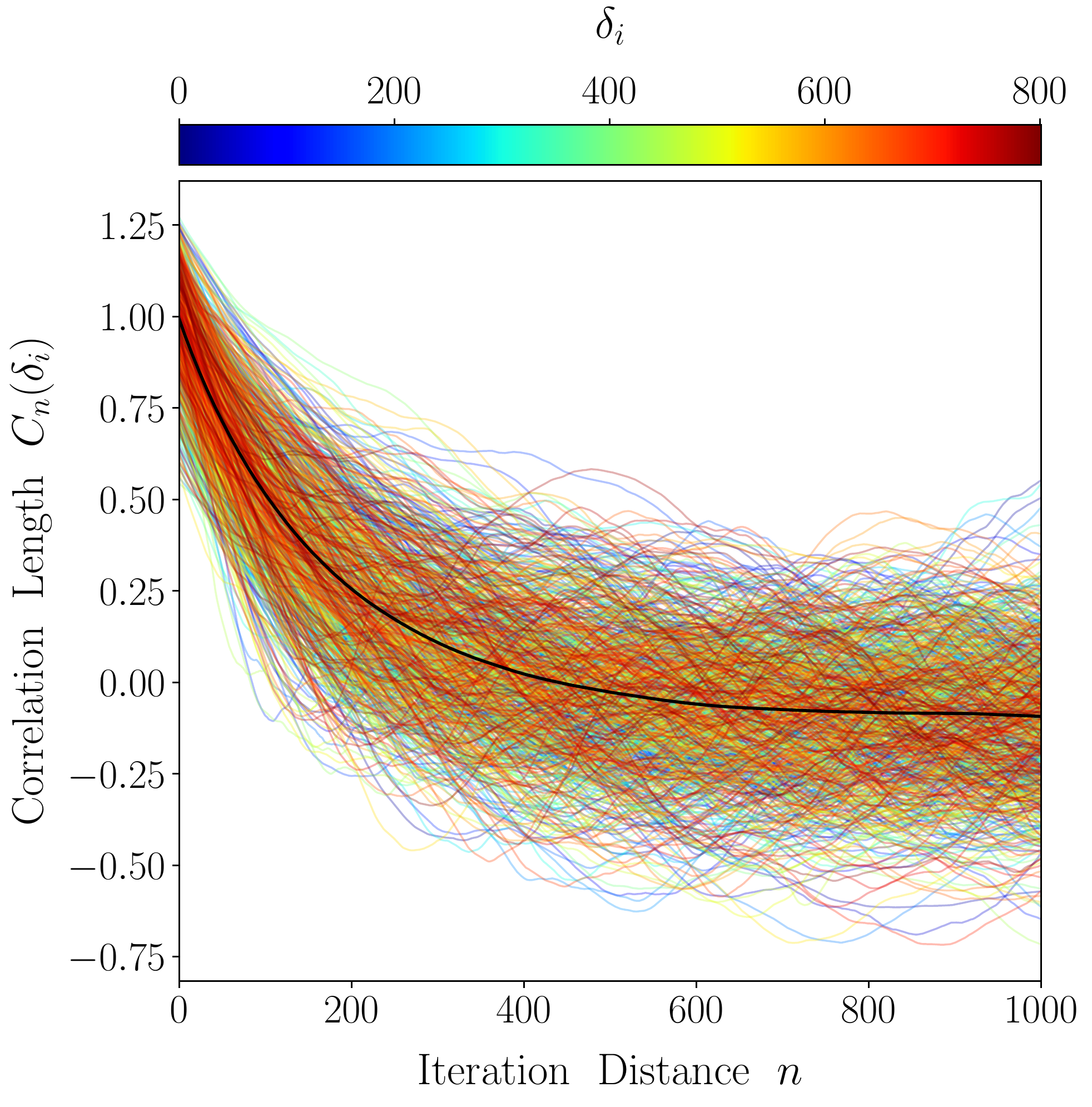}
\vspace{-0.05cm}\\
\includegraphics[width=7.2cm]{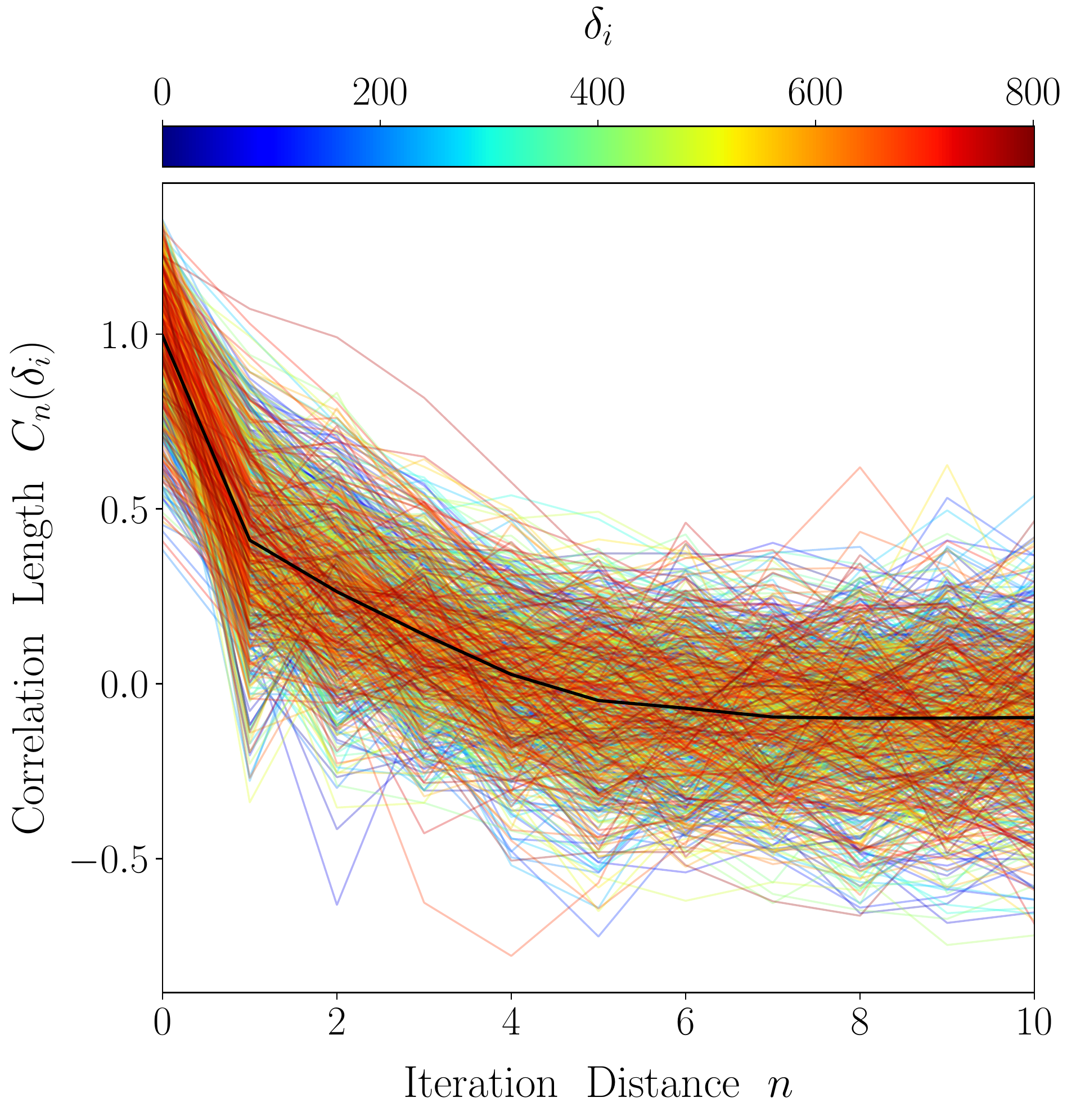}
\end{tabular}
\caption{\label{fig:corr}{\color{black} Correlation length for 800 randomly chosen density voxels $\delta_i$ as a function of the iteration distance. {\bf Upper panel:} $\mathcal{T}_4, i=3$, {\bf middle panel:} $\mathcal{T}_2, N_{\rm eval}=5$, and {\bf Lower panel:} $\mathcal{T}_2, N_{\rm eval}=80$. The black solid line represents the mean over all density fields. }} 
\label{fig:correlation}
\end{figure}

\subsubsection{Correlation length}
\label{sec:corr}

Finally, we compute the correlation length of the density bins over the iteration distance. The correlation length is calculated as
{
\ba
C_n(\sigma_{j})=\frac{1}{N-n}\sum_{i=0}^{N-n}\frac{\left(\delta_{j}^{i}-\langle \delta_{j}\rangle\right)\left(\delta_{j}^{i+n}-\langle\delta_{j} \rangle\right)}{\sigma^{2}(\delta_{j})}\,,
\ea
where $\delta_{j}$ is the overdensity field in each iteration, $N$ is the number of samples and $n$ is the distance between iterations.
Some particular computations are shown in figure \ref{fig:corr}.
}

{Table \ref{table_corr} shows  the correlation length for  different configurations previously studied (summarised in table \ref{table_conv}). In particular, it shows the correlation length in terms of evaluations of the Hamiltonian equations of motion and the improvement factor of each case in producing independent samples, given by the ratio between these evaluations and those ones of the reference scheme ($\mathcal{T}_4, i =3$). We also find the number of independent samples produced with $\mathcal{T}_4$, $i=3$ until each scheme converges 
\ba
{\rm NoIS}\equiv\frac{{\rm CONV}^{\rm eff}[{\rm scheme}]-{\rm CONV}^{\rm eff}[{\tau_4}_{i=3}]}{{\rm CL}^{\rm eff}[
{\tau_4}_{i=3}]}\,.
\ea
Then, we have the acceptance rate after the convergence, and the global step-size, also in this range, including the effective one. We can conclude from table \ref{table_corr} that, for the second-order leap-frog algorithm, the optimal configuration is the one of $N_{\rm eval}=80$. However, although this case has a very low correlation length of $4$ iterations, due to the high value of evaluations of the Hamilton's equations of motion in each iteration (we have an average of $40.5$ as we can see in the second column of the table), the fourth-order method is still more efficient, as we discuss in detail below, having an average correlation length of $18$ iterations. Another important aspect to consider here is that, while the case $\mathcal{T}_2$, $N_{\rm eval}=80$ converges, the $\mathcal{T}_4$, $i=3$ has already produced $\sim280$ independent samples. We can also see that, although for the convergence the case of $\mathcal{T}_2$,  $N_{\rm eval}=5$ was closely the optimal one, for the correlation length it is a factor $\sim5$ times worse than the fourth-order method.}

\begin{figure}
\centering
\includegraphics[width=8.5cm]{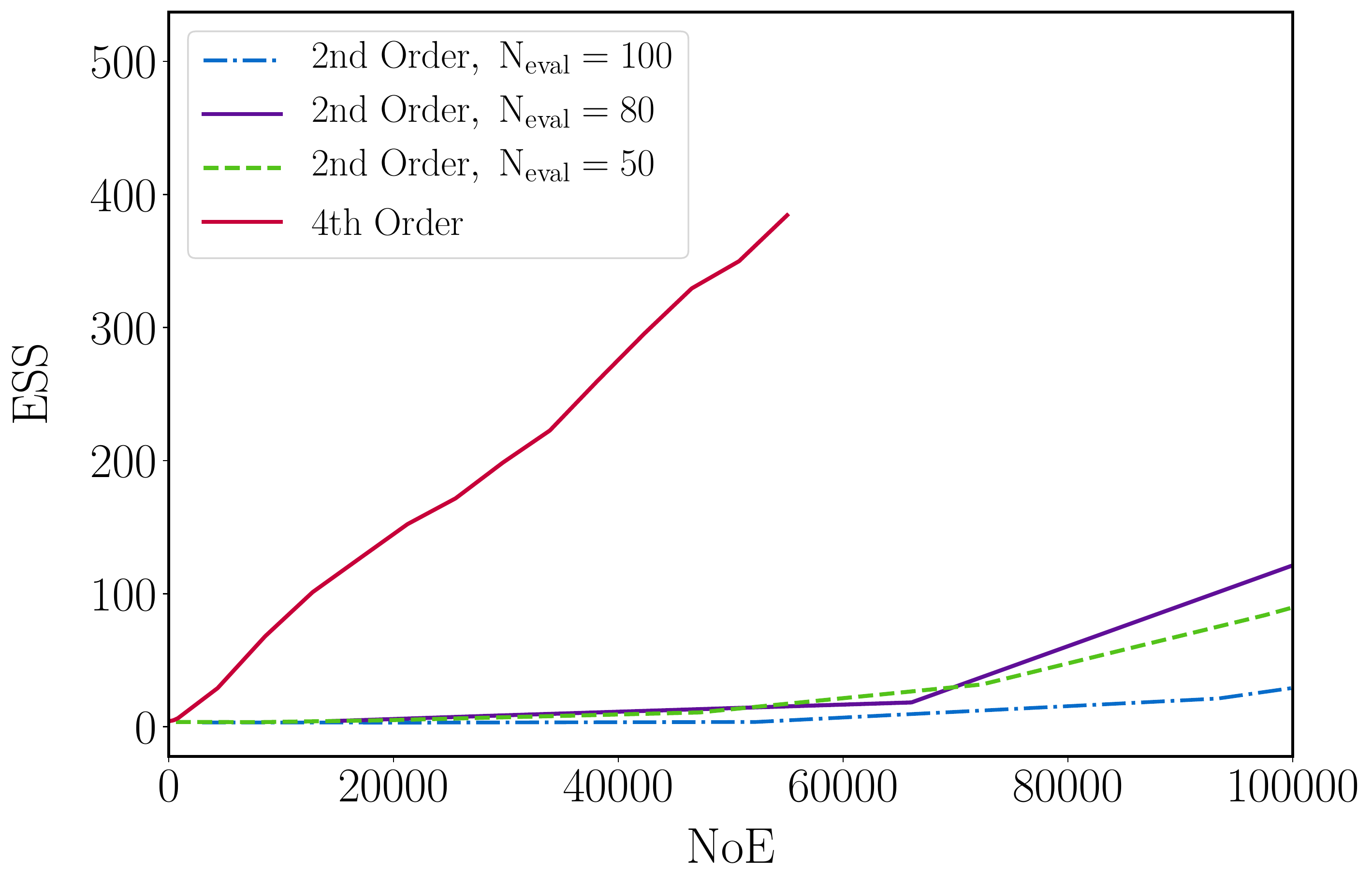}
\caption{\label{fig:ess-NoE} {\color{black} ESS estimator as a function of the number of evaluations of the Hamilton's equation of motion (NoE), for the fourth  and the second (with  $N_{\rm eval}= 50, 80, 100$) order schemes. 
}}
\end{figure}

\begin{figure*}
\centering
\includegraphics[width=17cm]{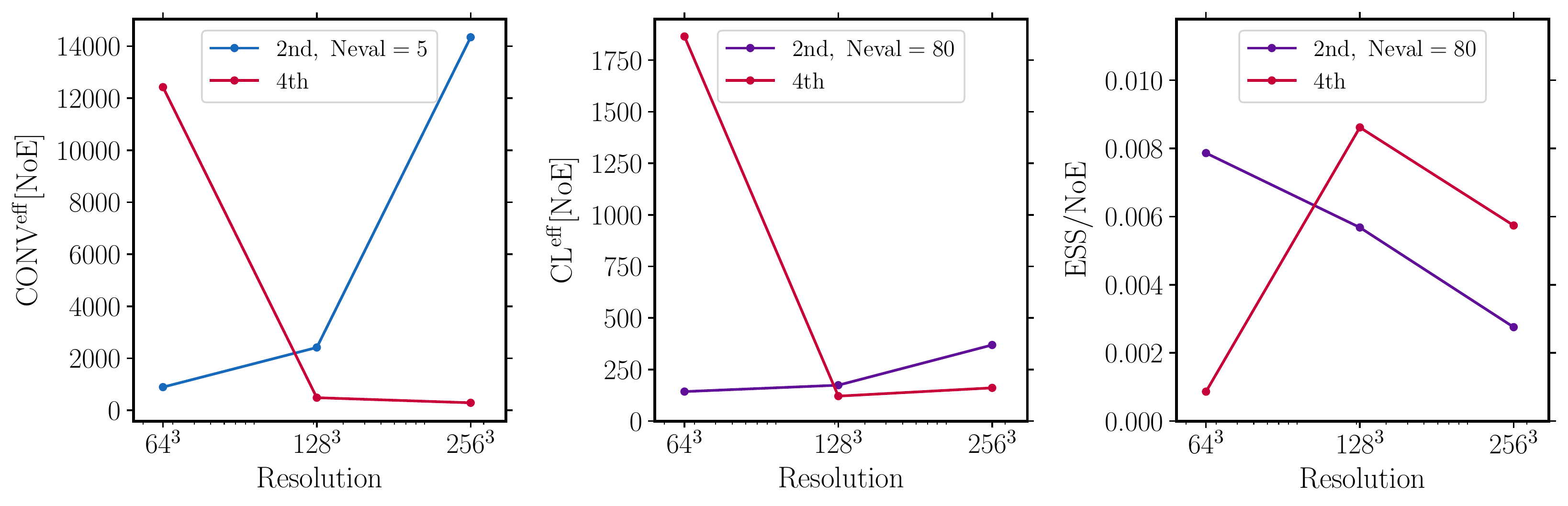}
\rotatebox{90}{\put(77,-185){$\langle$\hspace{1.3cm}$\rangle$}}
\caption{\label{fig:resolution} {\color{black} {\bf Left panel:} Average number of evaluations of Hamilton's equations of motion to achieve  convergence (CONV$^{\rm eff}$[NoE]) as a function of  resolution ($64^{3}$, $128^{3}$ and $ 256^{3}$ cells), with the red and the blue lines showing the fourth  and   second  ($N_{\rm eval}=5$) order schemes, respectively.  {\bf Middle panel:}  correlation length in terms of number of evaluations of  Hamilton's equation of motion, also as a function of  resolutions. The red and purple lines represent the fourth and  second ($N_{\rm eval}=80$) order scheme, respectively. {\bf Right panel:} averaged (from iterations 1000 to 4000) ESS estimator normalised by the number of evaluations of Hamilton's equations of motion (NoE) as a function of  resolution. The red and  purple lines represent the fourth   and  second  ($N_{\rm eval}=80$) order scheme, respectively. For each discretisation case, the most efficient set-up has been selected. The optimal basic $\epsilon$ step-sizes were found for the 128$^3$ and 256$^3$ cases. We did not do such a study for the second-order leap-frog 64$^3$ case, since it became already clear that it is superior.}} 
\end{figure*}

{\color{black} Figure \ref{fig:correlation}  shows the correlation length of for the different  leap-frog algorithms. The upper panel presents the best fourth-order case, while the middle and lower panels show the second-order cases for $N_{\rm eval}=5$ and $N_{\rm eval}=80$, respectively.
 The black solid line represents the mean over all density voxels $\delta_{i}$, where  we assume that we have independent samples, when the correlation length is lower than $0.1$. 
  We have defined an effective time step-size shown in equation   \ref{eq:effective_tau} trying to understand the trends seen in tables \ref{table_conv} and \ref{table_corr}.   All fourth-order schemes show larger effective time steps than the second-order ones.
 We also find that the  methods with the largest effective time steps converge faster (see table   \ref{table_conv}). 
  
  However, the shorter correlation length for the second-order case, achieved at $N_{\rm eval}=80$, has not the largest effective time step-size (table   \ref{table_corr}). 
  
   {\color{black} 
   \subsubsection{Effective correlation length}
   
  Larger integration times resulting from the product of the time step-size and the number of evaluations will produce less correlated samples.  However, there is a trade-off from increasing the number of evaluations, and at some point, the high $\langle{\rm NoE}\rangle$ combined with low acceptance rates do not compensate for the shorter CL [NoI]. Also, towards larger integration time steps, the acceptance rate diminishes. What counts at the end is the effective correlation length CL$
 ^{\rm eff}$.

 One needs to ask how many gradient evaluations NoEs are required to obtain the number of  accepted iterations given by CL[NoI]. Across a chain after convergence, this effective CL fluctuates depending on the rejections, which have to be included in the computation of the NoEs. Therefore, we take the average.
 Hence, the effective correlation length CL$^{\rm eff}$ is computed as the average number of evaluations (NoEs) including rejected samples, required to obtain the number of accepted iterations as indicated by the respective correlation length CL. The average is computed well after convergence, taking several thousands of iterations using different seeds.
We consider that this is a direct way of estimating the number of evaluations required to get independent posterior samples.
Nonetheless, there are some alternative ways of estimating this in the literature. 

\subsubsection{Effective  sample size}

The effective sample size (ESS) per posterior evaluation is defined as
 \ba
 {\rm ESS} (\{\delta_i\}_1^N) = \frac{N}{1+2\sum_{n=1}^{N-1}(1-\frac{n}{N})C_n}\, .
 \ea
Even if the correlation length $C_n$ in theory should asymptotically drop to 0 after a finite number of iterations $N$, in practice noise will dominate the ESS estimator for large sums. 
This is why many different solutions have been suggested in the literature \citep[see different definitions in][]{JSSv076i01,kaplan2014,2013PASP..125..306F}.
 We decide to follow \citet{JSSv076i01} and \citet{2017arXiv170607561S}
 to overcome these problems by truncating the sum  over the
correlation length when it goes below 0.1.
The results of this computation 
 are shown in figures \ref{fig:ess}-\ref{fig:resolution}.
This demonstrates an efficiency of the fourth-order sample over the second-order one of a factor of about 3.0, which is superior to our direct estimation of 2.3 with the CL$^{\rm eff}$.
While our direct CL$^{\rm eff}$ computation assumes a constant CL after convergence (see section \ref{sec:conv}), the ESS estimator does not. The advantage of the CL$^{\rm eff}$ over the ESS is only that it is very fast to compute, and qualitatively yields the same results, as can be seen in figure \ref{fig:resolution}.
But, in terms of statistical robustness we rather rely on the ESS estimator.

\begin{figure}
\begin{center}
\vspace{-0.cm}
\includegraphics[width=8cm]{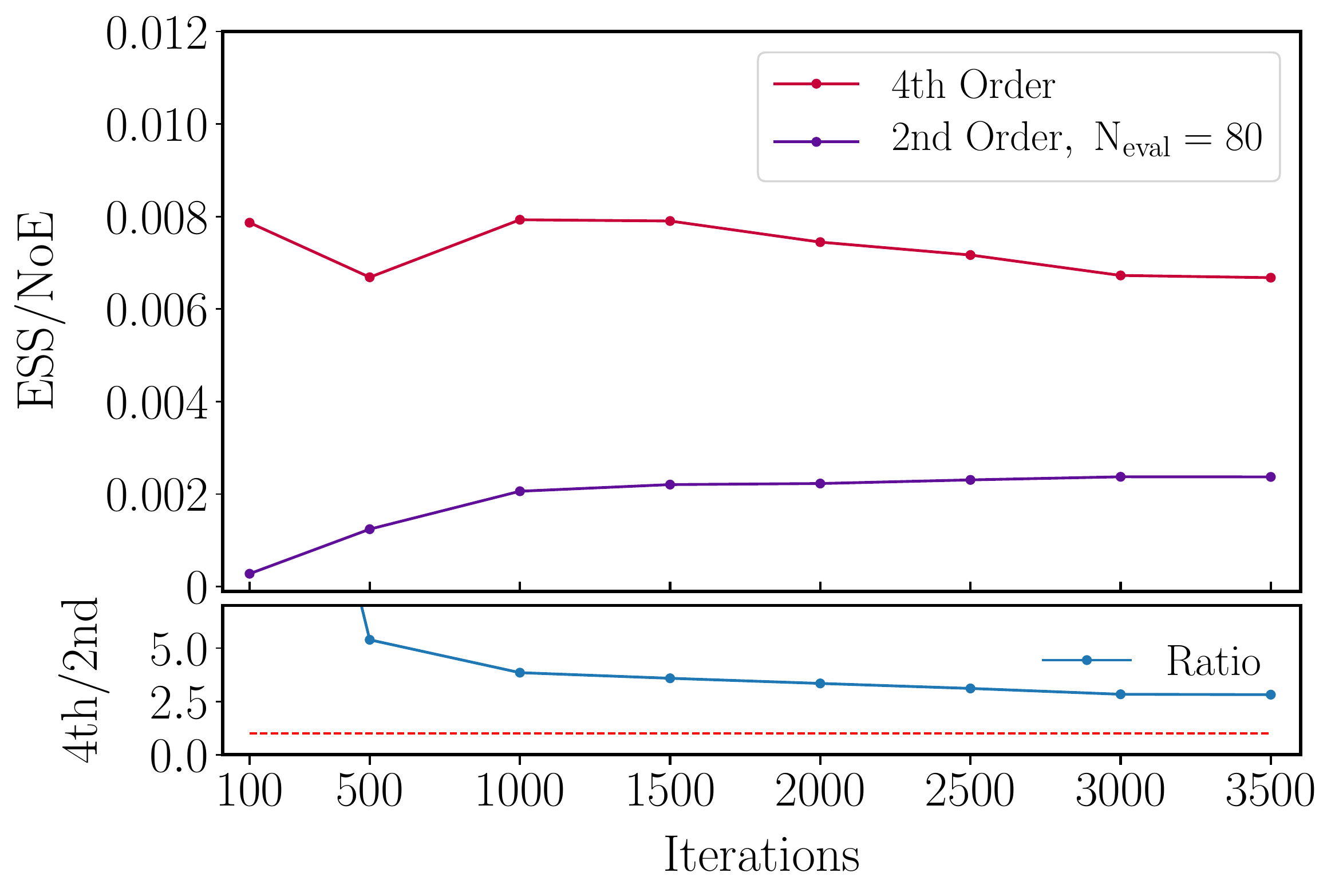}
\end{center}
\vspace{-0.cm}
\caption{{\color{black}ESS estimator divided by the number of evaluations of the Hamilton's equations of motion (NoE). {\color{black} The lower caption shows the ratio between the fourth and the second-order cases.} The fourth-order method is on average a factor 3.2 superior to second-order one, considering the range after 1000 iterations iterations (2.8 for the range 3000 to 3500 iterations).  We have checked that the rest of  $N_{\rm eval}$ cases (including $N_{\rm eval}=50$)  perform worse than $N_{\rm eval}=80$. }}
\label{fig:ess}
\end{figure}

}


}

    We can conclude from this study that the most efficient way to perform the second-order leap-frog algorithm is to start with a low number of evaluations ($N_{\rm eval}=5$) until it reaches convergence, and then change to a high number ($N_{\rm eval}=80$) to reduce the correlation length.  Our calculations also demonstrate that the fourth-order case delivers the most efficient posterior sampling calculations  {\color{black} for high statistical dimensions} cases.
    We have not fully studied the potential of the fourth-order scheme.
    More efficient fourth-order schemes might be  obtained by considering random  $i$ within  the range $2,\dots,4$, or other combinations. Such methods could be improved by also  considering different probabilities for each $i$ value.  This  study is out of the scope of this work.

{\color{black} 
\subsection{Dimensionality and efficiency}

 To confirm that higher order schemes become more important with increasing number of dimensions in parameter space, we make an additional resolution analysis.
This has the additional property of decreasing the uncertainty per dimension (i.e. per cell), since the number density of galaxies increases, and consequently the uncertainty in the matter field (the relative Poisson error) decreases. 

The results of this study are shown in figure \ref{fig:resolution}. 
We find that the second-order scheme becomes superior to the fourth-order scheme, going down in resolution to a mesh of  64$^3$ cells.
The fourth-order scheme starts to become moderately more efficient  considering meshes of 128$^3$ cells, and clearly superior for the 256$^3$ case.
This is shown in the left panel in the burn-in phase.
The middle and right panels show the consistency between the effective correlation function CL$^{\rm eff}$ and the averaged effective sample size normalised by the respective number of evaluations $\langle$ESS/NoE$\rangle$, respectively (after convergence).

Note, that 64$^3$   corresponds for a volume of 1250 $h^{-1}$ Mpc to a cell resolution of about  20 $h^{-1}$ Mpc. This is considered a too low resolution for most  practical cosmological reconstruction cases \citep[see, e.g.][]{2017MNRAS.467.2331V}.
Current galaxy surveys trace  larger cosmic volumes than the one considered in this study making a mesh of 64$^3$ far from being useful.  

By considering larger uncertainties per dimension (i.e. per cell), we can find  higher order schemes becoming more efficient than the second-order one. Such a situation corresponds to lowering the number density (by say an order of magnitude) in the halo catalogue, which implies getting a poorer representation of the underlying dark matter field.
However, this becomes in general unrealistic, as we have considered the most massive haloes hosting luminous red galaxies.
When considering higher number densities we might find fourth-order schemes to perform worse than the second-order one on meshes of 256$^3$ cells. However, we will be in general interested in covering very large volumes considerably increasing the mesh size. Also, our full reconstruction approach considers tracers in Lagrangian space, where they are more sparse \citep[][]{2019arXiv191100284K}.

In conclusion, it will depend on  the particular problem which scheme performs better, but with increasing cosmic volumes second-order schemes will,  in general,  become a worse choice.
 
  }

\section{Conclusions}

This work presents an efficient Hybrid Markov Chain Hamiltonian Monte Carlo Sampling method for cosmological large-scale structure analysis. In particular, it relies on a fourth-order symplectic integration of Hamilton's equations of motion.
This is achieved through an operator formalism in which the original leap-frog algorithm is recursively applied in a combination of two forward time integration steps with an intermediate backward step and appropriate step-sizes. 
{\color{black} One of the key ingredients is to realise that the higher integration accuracy of the fourth-order scheme permits one to fix the number of evaluations of the equations of motions to a few ($\sim$7), being able to perform larger effective time steps in each evaluation of Hamilton's equations of motion, and obtaining high acceptance rates. }
{\color{black} At low resolutions (i.e. low dimensional spaces) we find, however, that the second-order scheme is superior, confirming the theoretical expectations \citep[][]{2017arXiv171105337B}.} 
We have restricted this study to the lognormal-Poisson model, applied to a full volume halo catalogue in real space on a cubical  mesh of $1250$ $h^{-1}$ Mpc, with 128$^3$ and 256$^3$ cells. However, we have shown that selection effects, redshift space distortions, and displacements can be accounted for within a Gibbs-sampling scheme, as implemented in the  \texttt{COSMIC BIRTH} algorithm.
In this way, the scheme presented here permits one to efficiently sample the primordial density fluctuations of the Universe from galaxy surveys within a posterior Bayesian inference framework \citep{2019arXiv191100284K}. This scheme can help to improve the efficiency of other Bayesian inference methods \citep[e.g., the publicly available \texttt{BARCODE} ][]{2019MNRAS.488.2573B}.

We  have demonstrated performing an extensive parameter study, that going from the usual second to fourth-order in the discretisation of  Hamilton's equations of motion improves the convergence  by a factor of  {\color{black} $\sim 30$} in number of evaluations for the best second-order case.
{\color{black} 
This implies, that   75-90 independent samples  are  obtained,  while  the  fastest  second-order  method  converges. 
Moreover, we obtain  independent samples  about $\sim3.0$ times faster than the best second-order scheme, which has a different setting than for the burn-in phase}. 
It is interesting to note, that the most efficient fourth-order case in the burn-in phase is the same, as the one after convergence. This is very convenient, when the ideal set-up of a Hamiltonian sampler  for a particular case needs to be investigated. 
{\color{black} We leave further investigations of more sophisticated higher order schemes for future work (see \citet[][]{2002McLachlanQuispel} and \citet[][]{Blanes_2014}, which converges to a 4th order scheme in the limit of vanishing step sizes).}

In summary, the {\color{black} investigation of this work shows that improved symplectic integrators can play a major role in gaining  computational efficiency for Hamiltonian Monte Carlo sampling methods in high dimensional problems} to go towards a full Bayesian analysis of the cosmological large-scale structure for upcoming galaxy surveys.

{\color{black}

\section*{Data in this article}

The data and c++ codes used in this article will be shared on reasonable request to the corresponding author.  
The halo catalog from the BigMD simulation can be obtained at this site: \url{https://www.cosmosim.org/cms/simulations/bigmdpl/}.
The ESS estimator used in this study is publicly available at: \url{https://github.com/gmetin/MCMC/blob/master/corr_length_birth.ipynb}.
 A python version of the c++ code used in this study is public at: \url{https://github.com/pacoshu/HMC} making a comparison of the efficiency between the second and fourth-order schemes  straightforward.

}

\section*{Acknowledgments}

{\color{black}The authors thank Jes\'us Sanz-Serna for explaining them the mathematical reason for the higher efficiency of higher order schemes over the  traditional second-order leap-frog scheme towards high dimensions. We also want to thank the STAN team for useful discussions and their interest in higher order schemes although, as it has become clear, the majority of the applications does not require the  high dimensionality of the problem considered in this work. }
{\color{black}The authors thank the anonymous referee, Jorge Mart{\'i}n Camalich, Mattia Dallabrida,  Andr{\'e}s Balaguera-Antol{\'i}nez and Florent Leclerq   for useful comments.} MHS thanks the {\it Astrofísicos Residentes} grant at the IAC for permitting her to work on this study as part of her  master thesis presented in July 2018. FSK acknowledges financial support from the Spanish Ministry of Economy and Competitiveness (MINECO) under the Severo Ochoa program SEV-2015-0548, and for the grants RYC2015-18693 and AAYA2017-89891-P.  
MA thanks for the hospitality at the IAC and the support from the Kavli IPMU fellowship that permitted him to  develop the analysis codes for the ESS estimator, CL, and RG test, used in this work.
CDV acknowledges the support of the Ministry of Science, Innovation and Universities (MCIU) through grants RYC-2015-18078 and PGC2018-094975-B-C22.



\bibliographystyle{mnras}
\bibliography{pfcbib.bib}




\bsp	

\label{lastpage}
\end{document}